\pdfoutput=1
\documentclass[12pt,letterpaper]{article}
\usepackage{jheppub}
\usepackage{journals}
\usepackage{graphicx}
\usepackage{placeins}
\usepackage[arrow, matrix, curve]{xy}
\def\Nfour      {\mathcal{N}\,{=}\,4}
\def\Nc         {N_{\rm c}}

\begin          {document}
\title          {Studying a charged  quark gluon plasma via holography and higher derivative corrections}
\author[a]      {Sebastian~Waeber}
\author[a]{Andreas~Sch\"afer}
\affiliation[a] {Institute for Theoretical Physics, University of Regensburg,
                D-93040 Regensburg, Germany}
                \emailAdd       {sebastian.waeber@physik.uni-regensburg.de}
                \emailAdd       {andreas.schaefer@physik.uni-regensburg.de}
\abstract
    {%
  We compute  finite 't Hooft coupling corrections to observables related to  charged quantities in a strongly coupled  $\mathcal{N}=4$
    supersymmetric Yang-Mills plasma. The coupling corrected equations of motion of gauge fields are explicitly derived and differ from findings of previous works, which contained several small errors with  large impact. As a consequence the $\mathcal{O}(\gamma)$-corrections to the
    observables considered, including the
    conductivity, quasinormal mode frequencies, in and off equilibrium spectral density and photoemission rates, become much smaller. This suggests that infinite coupling results obtained within AdS/CFT are little modified for the real QCD coupling strength. 
   
    }
\maketitle
\section{Introduction}
Experimental data from heavy ion collisions at LHC and RHIC suggest that the produced quark gluon plasma (QGP) is strongly coupled and equilibrates extremely fast. Unfortunately, standard QCD techniques are unsuitable to treat the strongly coupled, non-equilibrium early dynamics. Therefore the best known way to study the early phases of the QGP before thermalization happened is via holography, by mapping weakly coupled supergravity (SUGRA) to its strongly coupled quantum field theoretical dual. Although there is no dual description for QCD one can approach the real world by studying the plasma with the help of the  holographic dual of large-$N$, $\mathcal{N}=4$ strongly coupled super Yang-Mills (SYM) theory.  \newline \indent The QGP produced during heavy ion collisions lies somewhere in between the two extreme limits of infinitely strong coupling (or small curvature) with 't Hooft coupling $\lambda=\infty$  and weak coupling, which would allow for a perturbative description. One way of investigating this region is to consider finite coupling corrections or higher derivative corrections to the type IIb SUGRA action. These additional contributions of order $\mathcal{O}(\alpha'^3)$ for the dual gravity theory, where $\alpha'$ is related to the string length $l_s$ via $\alpha'=l_s^2$, yield finite coupling corrected correlators, emission rates, transport coefficients etc. on the QFT side.  \newline \indent  One interesting topic in this context is the analysis of the behaviour of   charged particles in such a QGP. In recent years there have been several works contributing to a deeper quantitative understanding thereof. One important step was the computation of leading  coupling corrections to the equations of motion  of gauge fields in a strongly-coupled $\mathcal{N} = 4$ SYM plasma by considering $\mathcal{O}(\alpha'^3)$ corrections to the type IIB  supergravity action ~\cite{Hassanain:2011fn, Hassanain:2012uj}. These $\alpha'$-corrected equations of motion were then used to study the conductivity,  the transport coefficient in this channel and the photoemmission rate, which give important information about the structure of the plasma. Determining $\alpha'$ corrections to these quantities is of  major interest, especially since this allows first cautious comparisons and interpolations between the spectra of strongly coupled and weakly coupled plasmas ~\cite{Hassanain:2012uj}.
Unfortunately the authors of  ~\cite{Hassanain:2011fn, Hassanain:2012uj} used a $5$-form that didn't solve its higher derivative corrected EoM. In addition, unlike stated in these papers,  the calculation was done in Euclidean signature, but the five form wasn't transformed appropriately. More specifically, we can reproduce their results, if we leave out an actually needed factor $i$ in front of the five form components of the form $dt \wedge \dots$ after the transformation to Euclidean signature. Also several terms contributing to  the Hodge duals got lost. Our first aim is to give a corrected derivation of the higher derivative corrected EoM for gauge fields in type IIb SUGRA. After that we revisit the computation of several  observables, whose $\alpha'^3$-corrections so far have been calculated with the EoM form  ~\cite{Hassanain:2011fn, Hassanain:2012uj}. In general we find that the actual higher derivative corrections to all quantities studied in this paper turn out to be substantially smaller than the values found in the literature so far. For instance in   ~\cite{Hassanain:2011fn}
the correction factor to the conductivity was given as $(1+\frac{14993}{9}\gamma)$, whereas we obtained  $(1+125 \gamma)$. A comparison with the transport coefficient of the spin $2$ channel is given in table \ref{table:1}.
In contrast to previous works we find that the behaviour of the  photoemission rate  and spectral density at finite coupling agree with expectations from weak coupling calculations in both the small and, that is new, the large energy limit \cite{CaronHuot:2006te}. In \cite{CaronHuot:2006te} the authors derived that in the weak coupling limit decreasing coupling means increasing phtotoemission rate at small momenta and decreasing photoemission rate  at large momenta. The signs of the correction factors we found coincide with these expectations.  We start from the higher derivative corrected type IIb action and compute finite coupling corrected QNM spectra, spectral density, photoemission rate and conductivity of the plasma. Before we come to finite coupling corrections we give a detailed description how to introduce gauge fields in type IIb SUGRA by twisting the five sphere along certain angles, which was first described in \cite{Chamblin} .
We try to provide enough details of the calculations to allow the reader to check it with limited effort.
\section{Einstein-Maxwell-AdS/CFT in the $\lambda \to \infty$ limit } 
\begin{figure}
\begin{equation*}
\begin{xy}
  \xymatrix{
      (C_4)_{x y_2 y_4 y_5} \ar[r]^d \ar[rd]^d \ar[rdd]^d \ar[rddd]^d   &   (F_5)_{t x y_2 y_4 y_5}\ar[r]^{*} &   (*F_5)_{u y z y_ 1y_3}\ar[r]^d \ar[rdd]^d   & (d*F_5)_{t u y z y_ 1y_3} \\
         & (F_5)_{ x z y_2 y_4 y_5}\ar[r]^{*} &   (*F_5)_{tu y y_ 1y_3} \ar[ru]^d \ar[rddddddd]^d  & \\
      (C_4)_{x y_1 y_4 y_5}  \ar[rd]^d \ar[rdd]^d \ar[rddd]^d \ar[rdddd]^d   &   (F_5)_{u x y_2 y_4 y_5}\ar[r]^{*} &   (*F_5)_{t  y z y_ 1y_3}\ar[ruu]^d \ar[rdddd]^d & (d*F_5)_{ u y z y_ 1 y_2y_3} \\    
     & (F_5)_{ x  y_1 y_2 y_4 y_5}\ar[r]^{*} &   (*F_5)_{tu  y z y_3} \ar[ruuu]^d \ar[rd]^d & \\
       (C_4)_{t x y_4 y_5} \ar[ruuuu]^d \ar[r]^d \ar[rdddd]^d \ar[rddd]^d   &   (F_5)_{t x y_1 y_4 y_5}\ar[r]^{*} &   (*F_5)_{u y z y_ 2y_3}\ar[r]^d \ar[ruu]^d  & (d*F_5)_{t u y z y_ 2y_3} \\
       & (F_5)_{  u x y_1 y_4 y_5}\ar[r]^{*} &   (*F_5)_{t y z y_2y_3} \ar[ru]^d \ar[rd]^d  & \\
    (C_4)_{x z  y_4 y_5} \ar[ruuuuu]^d\ar[r]^d\ar[rd]^d \ar[rddd]^d &   (F_5)_{ x z y_1 y_4 y_5}\ar[r]^{*} &   (*F_5)_{t u y  y_2y_3}\ar[ruu]^d \ar[rdd]^d   & (d*F_5)_{t  y z y_ 1 y_2y_3} \\
      & (F_5)_{u x z y_4 y_5}\ar[r]^{*} &   (*F_5)_{u y y_ 1 y_2 y_3}\ar[rd]^d \ar[ruuuuu]^d  & \\
(C_4)_{u x  y_4 y_5}\ar[rd]^d  \ar[r]^d\ar[ruuu]^d \ar[ruuuuuu]^d &   (F_5)_{t u x y_4 y_5}\ar[r]^{*} &   (*F_5)_{ y z y_ 1y_2 y_3}\ar[ruu]^d \ar[ruuuuuu]^d  & (d*F_5)_{tu y y_1 y_2 y_3} \\
    & (F_5)_{u x z y_4 y_5}\ar[r]^{*} &   (*F_5)_{ty  y_ 1y_2y_3} \ar[ru]^d\ar[ruuu]^d.  & \\
  }
\end{xy}
\end{equation*}
\caption{Graphic depiction of the "closed" system of differential equations around the $x y_2 y_4 y_5$-direction of $C_4$. In this order in $\alpha'$ the right hand side should give zero.}
  \label{diag1}
\end{figure}
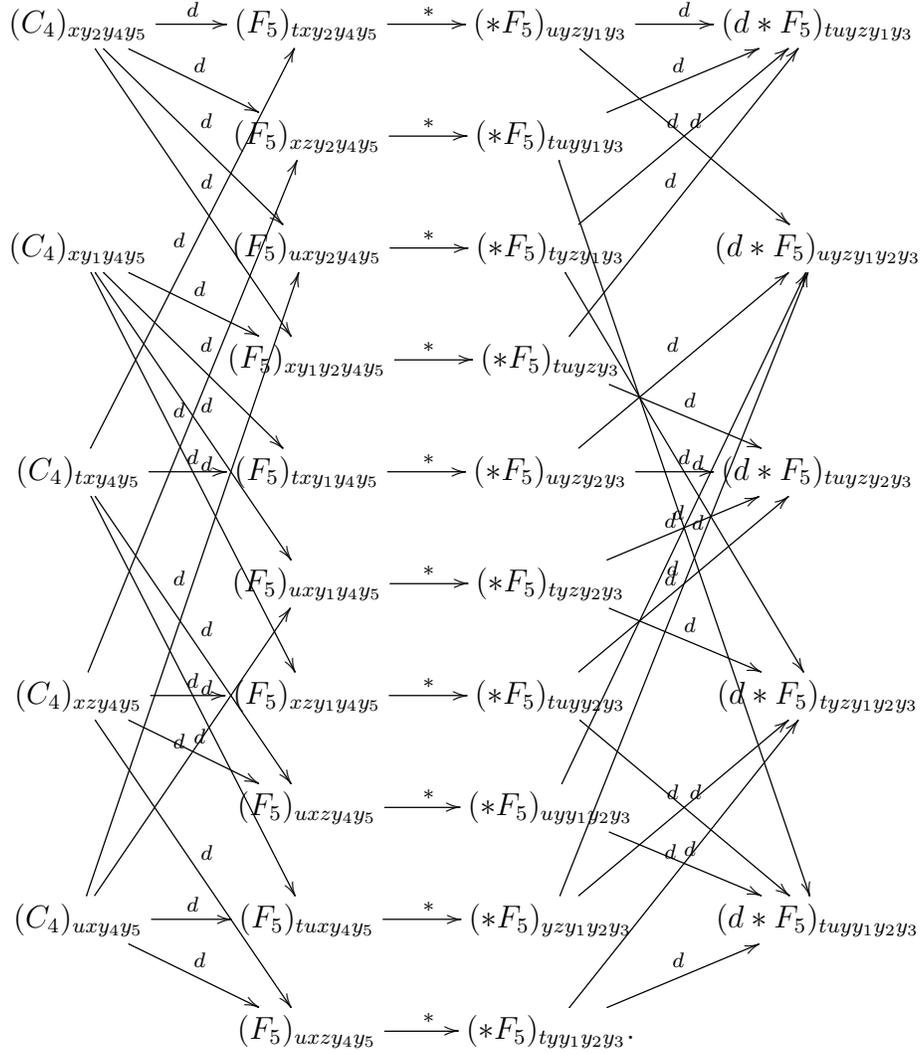

    The aim  of this section is to give a detailed description of  how to introduce charge and gauge fields in AdS/CFT starting from the type IIb  SUGRA action 
\begin{equation}
S_{10} = \frac{1}{2 \kappa_{10}} \int d^{10}x \sqrt{-\det(g_{10})} \bigg[R_{10}-\partial_\mu \phi \partial^\mu \phi-\frac{1}{4 \times 5!} F_5^2 \bigg],
\label{eq:action}
\end{equation}    
where $F_5$ is the $5$-form and $g_{10}$ the metric of the $10$ dimensional manifold.  In the following calculations we set the constant $l$, which measures the size of $S_5$, to $1$, since the resulting EoM for gauge fields won't depend on it.
   In   \cite{Chamblin} it was shown that in order to obtain Maxwell-terms $F_{\mu \nu}F^{\mu \nu}$ in the reduced $5-$dimensional theory one has to twist the five sphere $S_5$ along its fibers in a maximally symmetric manner. The  ansatz for the metric in this case  has the form
  \begin{equation}
  ds_{10}^2= ds_{\text{AdS}}^2+\sum_{i=1}^3 \big(d\mu_i^2+\mu_i^2 (d\phi_i+\frac{2}{\sqrt{3}} A_\mu dx^\mu)^2 \big),
  \label{metricc}
  \end{equation}
  with 
  \begin{align}
ds_{AdS}^2= &-r_h^2 \frac{1-u^2}{u}dt^2+ \frac{1}{4 u^2(1-u^2)} du^2+ \frac{r_h^2 }{u}(dx^2+dy^2+dz^2),
\label{metric}
\end{align}
where the unperturbed metric is just the AdS Schwarzschild black hole solution times $S_5$ with horizon radius $r_h$. It is convenient to work here with the following $S_5$-coordinates, for which we define $ \mu_i$ with $i \in \{1,2,3\}$ to be the direction cosines 
\begin{equation}
\mu_1= \sin(y_1), \hspace{1cm} \mu_2= \sin(y_2)\cos(y_1),\hspace{1cm} \mu_3= \cos(y_1) \cos(y_2)  ,
\end{equation} 
and set the angles
\begin{equation}
\phi_1= y_3, \hspace{1cm} \phi_2= y_4,\hspace{1cm} \phi_3= y_5,
\end{equation}
such that the metric of the $5-$sphere is given as 
  \begin{align}
d\Omega_5^2 =& \sum_{i=1}^3 \bigg(d\mu_i^2+\mu_i^2 d\phi_i^2 \bigg)=
 dy_1^2+\cos(y_1)^2dy_2^2+\sin(y_1)^2dy_3^2+\nonumber \\&\cos(y_1)^2\sin(y_2)^2
dy_4^2+\cos(y_1)^2\cos(y_2)^2dy_5^2.
\end{align}
It is straightforward to check that with this metric ansatz we obtain 
\begin{equation}
R_{10} = R_{10}^{A_\mu \to 0}-\frac{1}{3} F_{\mu \nu }F^{\mu \nu},
\end{equation}
with $F=dA$.
The dilaton  part of the action can be ignored here, since its EoM does not couple with those of $A_\mu$ and the solution of its EoM in this order in $\alpha'$ is simply zero. On the other hand it is crucial to understand in  detail the role of the five form part of the action in this calculation. In  the following we will  motivate its ansatz, which was given in \cite{Chamblin}. \\ \indent
The five form $F_5$ is not an independent field  with respect to which we have to vary the action in order to complete the set of  EoM for type IIb fields relevant in this case. Actually, the term $F_5^2$ in the action is the kinetic term of the $4$-form $C_4$ with $dC_4=F_5$, which straightforwardly leads to the EoM obtained by varying $\mathcal{S}_{10}$ with respect to $C_4$:
\begin{equation}
d * F_5=0,
\label{eqF51}
\end{equation} where $*$ is the Hodge star operator. In addition one  has $d F_5=0$, which already reveals the self dual structure of the solution for $F_5$ in this order  in $\alpha'$.\\
 \indent In the case of a vanishing gauge field $A_\mu=0$ the self dual solution to (\ref{eqF51}) is 
 \begin{equation}
 F_5^{\text{el}}= -4\epsilon_{\text{AdS}}=-4\sqrt{-g_{\text{AdS}}}dt \wedge du \wedge dx  \wedge dy \wedge dz,
 \label{sol1}
 \end{equation}
 
  \begin{equation}
 F_5= (1+*) F_5^{\text{el}},
 \label{sol2}
 \end{equation}
 where $\epsilon_{\text{AdS}}$ is the volume form of the $AdS$-part of the manifold. The forefactor $-4$ is chosen in such a way that in the dimensionally reduced action we have 
 \begin{equation}
\frac{\text{vol}(S_5)}{2\kappa_{10}}\int d^5x\sqrt{-\det(g_{AdS})}\bigg[R_5 -8 +R_{S_5}\bigg]=\frac{\text{vol}(S_5)}{2\kappa_{10}}\int d^5x\sqrt{-\det(g_{AdS})}\bigg[R_5 +12\bigg].
\end{equation}
\indent Now we want to find a solution for $d F_5=0$ and $d* F_5=0$ with the metric  (\ref{metricc}). In order to see that  $ F_5^{\text{el}}= -\frac{4}{l}\epsilon_{\text{AdS}}$ 
is no  longer  the correct ansatz we consider the $tuyzy_1y_3$-direction of the $6$-form $d* F_5$. In the following we only consider transverse fields, which means that only $A_x$ is non-vanishing and $A_x =A_x(u,t,z)$. The deduction for longitudinal fields is analogous.
Remember that we are interested in linearized differential equations for $A_\mu$, which we consider as tiny fluctuations of our background geometry. This means that terms of order $A_\mu^2$ or higher can be discarded, such that there are only $6$ non-diagonal elements in the matrix representation of the metric tensor $g^{\mu \nu}$, namely $g^{x y_3},g^{x y_4},g^{x y_5}$ and interchanges of $x$ and $y_i$. From our solution in the $A_\mu=0$ case we already know that we will at least have one non vanishing term in the $tuyzy_3$-direction of the $5$-form $* F_5$, which is proportional to
\begin{equation}
\sqrt{-g}g^{y_1 y_1}g^{y_2 y_2}g^{y_3 x}g^{y_4 y_4}g^{y_5 y_5}( F_5^{A_\mu \to 0})_{y_1 y_2 y_3 y_4 y_5}.
\end{equation}
Note that we are not making use of the sum-convention here and henceforth.
 This term is proportional to $A_\mu$ without any derivatives and has a non trivial $y_1$-dependence, such that we have
\begin{equation}
0 \neq(d* F_5)_{tuyzy_1y_3}=\partial_{y_1}(\sqrt{-g}g^{y_1 y_1}g^{y_2 y_2}g^{y_3 x}g^{y_4 y_4}g^{y_5 y_5}( F_5^{A_\mu \to 0})_{y_1 y_2 y_3 y_4 y_5})+\dots
\label{super}
\end{equation}
without further directions of $F_5$ being non zero.
This term can't be canceled by the EoM for $A_\mu$, since it would give a mass to our gauge field. Consequently  there have to be more components of the solution for $F_5$, which give non-zero contributions, such that these mass terms cancel. The symmetries of this problem should dictate, which directions of the five form vanish and which don't. We instead use a different approach. We start from the fact, that our final ansatz for the $C_4$ can  only depend on the coordinates $u,t,z,y_1,y_2$, i.e. the coordinates the metric and its fluctuations $A_\mu$ depend on. Any other  dependence would lead to non-vanishing components of $d*d C_4$. This means the only possible components of $C_4$ proportional to $A_\mu$ that could give a contribution to the $tuyzy_1y_3$-component of  $d*d C_4$ are $(C_4)_{x y_1 y_4 y_5},(C_4)_{x y_2 y_4 y_5},(C_4)_{xz y_4 y_5},(C_4)_{t x  y_4 y_5},(C_4)_{u x y_4 y_5}$ modulo permutations of their $4$ indices. In the following, when we address properties of certain directions of forms, e.g. for $(C_4)_{abcd}$ the $abcd$-direction of $C_4$, these properties' applicabilities implicitly include all permutations of the indices $a b c d$ with the correct signs.\\ \indent
Graphically we can depict all relevant contributions of these $4$-form components to the differential equations shortly written as $d * d C_4=0$ as shown in figure \ref{diag1}.
Note that this diagram is closed in the sense that plus the contribution in (\ref{super}) all terms contributing to the  $ t u y z y_ 1y_3$, $ u y z y_1 y_2 y_3$ , $ t u y z  y_2 y_3$, $ t y z y_1 y_2 y_3$ and $t u y  y_1 y_2 y_3$-directions of $d * F_5$ are depicted  and $(C_4)_{x y_1 y_4 y_5}$, $(C_4)_{x y_2 y_4 y_5}$, $(C_4)_{xz y_4 y_5}$, $(C_4)_{t x  y_4 y_5}$, $(C_4)_{u x y_4 y_5}$ do not contribute to any other directions of $d * F_5$. The next important observation is that $(d*F_5)_{ u y z y_ 1 y_2y_3} $, $(d*F_5)_{t  y z y_ 1 y_2y_3} $ and $(d*F_5)_{tu y y_1 y_2 y_3}$ cannot be set to 0 by imposing the EoM of $A_x$, because they contain
odd derivatives in the $t$ and $z$ direction $ \partial_zA_x,\partial_tA_x$ or $ \partial_z^3 A_x,\partial_t^3 A_x$, if we have only even derivatives in $(d*F_5)_{tu y z y_1 y_3}$. 
From the requirement that there are no mass terms in the EoM for $A_x$ we can deduce from (\ref{super}) and the form of $F_5^{A_\mu \to0}$ that $(*F_5)_{tuyzy_3}$ is proportional to $\sin(y_1)^2$ and has no $y_2$-dependence. Therefore, $(C_4)_{x y_1 y_4 y_5}$ doesn't contribute to $(d*F_5)_{tuyzy_1y_3}$ and $(C_4)_{x y_2 y_4 y_5}$ doesn't contribute to $(d*F_5)_{tuyzy_2y_3}$. Thus, it is legal to choose $(C_4)_{x y_1 y_4 y_5}=0$.
This leads to the beautiful result that in diagram \ref{diag1} the contributions of $(C_4)_{xz y_4 y_5}$, $(C_4)_{t x  y_4 y_5}$, $(C_4)_{u x y_4 y_5}$ to  $(d*F_5)_{t u y z y_ 1y_3}$ have the same form as those of $ (C_4)_{x y_2 y_4 y_5}$ and are indistinguishable in the  final EoM $(d*F_5)_{t u y z y_ 2y_3}=0$ , which means it is a legitimate ansatz to set them to $0$ and solve $(d*F_5)_{t u y z y_ 2y_3}=0$ for $(C_4)_{x y_2 y_4 y_5}$. This process has to be repeated for  two further cases (remember that we only considered the off diagonal element $g^{x y_3}$ so far), which together with the self duality of the $5$-form leads to the result
\begin{equation}
(F_5^0)^{el}=-4 \epsilon_{AdS}, \hspace{0.5 cm} (F_5^1)^{el}=\frac{1}{\sqrt{3}}\sum_{i=1}^3 d(\mu_i^2) \wedge d\phi_i \wedge \bar{*}F_2,
\end{equation}
and 
\begin{equation}
F_5=(1+*)((F_5^0)^{el}+(F_5^1)^{el}),
\end{equation} 
with $F_2=d A$.
Of course, it isn't a coincidence that the electric part of $F_5$ is proportional to $\mathcal{J}\wedge \bar{*}d A$, with the K\"ahler-form of the five sphere $\mathcal{J}$, and there are more and easier ways to deduce this five form solution. Since we will have little choice but to work with similar brute force in the $\mathcal{O}(\alpha'^3)$-case, due to the complexity of the higher derivative correction terms to the type IIb action, it is a good exercise to  already do this in the lowest order in $\alpha'$.
Notice that the requirement that we are allowed to make the  ansatz (\ref{metricc}) implies that the EoM for $A_\mu$ can be  obtained both by varying the action with respect to $A_\mu$ and from the $t u y z y_ 1y_3$, $t u y z y_ 2y_4$,$t u y z y_ 2y_5$,$t u y z y_ 1y_5$ and $t u y z y_ 1y_4$-directions of $d*dC_4=0$, simply by starting from the fact that the metric tensor $g^{\mu \nu}$  has only certain off-diagonal elements. Varying the action with respect to $A_\mu$ leads to the following well known EoM for transverse fields in order $\mathcal{O}(\gamma^0)$
\begin{equation}
\partial_u^2 A_x-\frac{2u}{1-u^2}\partial_u A_x+\frac{\hat{\omega}^2-\hat{q}^2(1-u^2)}{u (1-u^2)^2}A_x=0
\label{eq0}
\end{equation}
with $\hat{x}=\frac{x}{2 r_h}=\frac{x}{2 \pi T}$ for $x \in \{q,\omega \}$ and the horizon radius $r_h$.
 Before we address higher derivative corrections it is advisable to look in detail at the following calculational prescription of SUGRA to obtain an effective action solely for the metric: "Take the ansatz of the $5-$form, plug it back into the action and only consider the magnetic part of your $F_5$ and double its contribution, then vary with respect to the metric.". In order to be able to decide, whether we are allowed to make use of this, if we include higher derivative corrections, we must understand where this prescription comes from.
In the easiest case, where we do not consider $\alpha'$-corrections or gauge fields $A_\mu$, our solution for the five form is given in (\ref{sol1}), (\ref{sol2}). If we want to derive the EoM for general metric components from the type IIb action (\ref{eq:action}) we, of course, are not allowed to impose a dependence of the five  form on $g^{\mu \nu}$ on the level of the action. Instead we have to vary the five form part of the action as follows
\begin{align}
&\delta  \int d^{10}x \sqrt{-g} \bigg[-\frac{1}{4 \cdot 5!} F_5^2 \bigg]=-\frac{1}{4 }\delta  \int d^{10}x \sqrt{-g}\bigg[ g^{tt}g^{uu}g^{xx}g^{yy}g^{zz}(F_5^{el})_{tuxyz}^2 +\nonumber \\ & g^{y_1y_1}g^{y_2y_2}g^{y_3y_3}g^{y_4y_4}g^{y_5y_5}(F_5^{mag})_{y_1 y_2y_3y_4y_5}^2\bigg]=-\frac{1}{4 }\delta  \int d^{10}x \bigg[-
\sqrt{\frac{g_{y_1 y_1}g_{y_2 y_2}g_{y_3 y_3}g_{y_4 y_4}g_{y_5 y_5}}{g_{tt}g_{uu}g_{xx}g_{yy}g_{zz}}}
 \nonumber \\ & (F_5^{el})_{tuxyz}^2 +\sqrt{\frac{g_{tt}g_{uu}g_{xx}g_{yy}g_{zz}}{g_{y_1 y_1}g_{y_2 y_2}g_{y_3 y_3}g_{y_4 y_4}g_{y_5 y_5}}}(F_5^{mag})_{y_1 y_2y_3y_4y_5}^2 \bigg],
\end{align}
which leads to a contribution to the EoM for $g^{\mu \nu}$ of the form
\begin{align}
4 \bigg((-1)^{1+\sum_{i=1}^5\delta_{\mu y_i}}\frac{\sqrt{-g}}{2} g^{\mu \nu}-(-1)^{\sum_{i=1}^5\delta_{\mu y_i}}\frac{\sqrt{-g}}{2} g^{\mu \nu}\bigg).
\end{align}
The same result is obtained from plugging the solution of the five form back into the action, only considering the contribution of the magnetic part times $2$.  This calculation can be performed similarly for more complicated five form solutions involving gauge fields. This recipe, which   is nothing but a calculational tool, is equivalent to the more intuitive but also more tedious approach of treating every metric component and every 4-form component as an independent field on the level of the action, varying with respect to all of them and solving the resulting system of EoM.  One important lesson to learn here is that the justification for this  prescription requires a self dual five form and we will see in the next section, that self duality is violated when we include higher derivative corrections (also see \cite{Paulos:2008tn}). We don't want to imply that this prescription breaks down for all non self dual forms, but  we are not aware of a justification to use it   to deduce the EoM for higher orders in $\alpha'$. Out of caution we will avoid this simplification in order $\mathcal{O}(\alpha'^3)$ and strictly following the variational principle.
\FloatBarrier
\section{Finite coupling corrections to the EoMs of   gauge fields}
\label{sec3}
\begin{figure}
 \begin{equation*}
\begin{xy}
  \xymatrix{
      (C_4)_{t y z y_3} \ar[r] \ar[rd] \ar[rdd]   &   (F_5)_{t u y z y_3}\ar[r]^{*} &   (*F_5)_{x y_1 y_2 y_4 y_5}\ar[r] \ar[rd] \ar[rdddddddd]   & (d*F_5)_{u x y_1 y_2 y_4 y_5} \\
      (C_4)_{t u y y_3}  \ar[ru] \ar[rddd] \ar[rdd]     & (F_5)_{ t y z y_1 y_3}\ar[r]^{*} &   (*F_5)_{u x y_2 y_4 y_5}  \ar[ru] \ar[rd] \ar[rdd]  & (d*F_5)_{t x y_1 y_2 y_4 y_5}  \\
      (C_4)_{u y z y_3}  \ar[ruu] \ar[rddd] \ar[rdddd]   &   (F_5)_{t y z y_2 y_3}\ar[r]^{*} &   (*F_5)_{u x y_1 y_4y_5} \ar[ruu] \ar[rdd] \ar[rddddd] & (d*F_5)_{u x z y_1 y_4y_5} \\    
    (C_4)_{y z y_1 y_3}  \ar[ruu] \ar[rdd] \ar[rdddd]    & (F_5)_{t u y y_1 y_3}\ar[r]^{*} &   (*F_5)_{x z y_2 y_4 y_5}  \ar[rddddd] \ar[ru] \ar[rdd] & (d*F_5)_{t u x y_2 y_4y_5} \\
       (C_4)_{t y y_1 y_3}  \ar[ruuu] \ar[ru] \ar[rdddd]   &   (F_5)_{t u y y_2 y_3}\ar[r]^{*} &   (*F_5)_{x z y_1 y_4 y_5} \ar[rdd] \ar[rddd] \ar[rdddd]  & (d*F_5)_{t u x y_1 y_4 y_5} \\
     (C_4)_{t y y_2 y_3}  \ar[rddd] \ar[ruuu] \ar[ru]    &   (F_5)_{u y z y_1 y_3}\ar[r]^{*} &   (*F_5)_{t x y_2 y_4 y_5} \ar[r] \ar[ruu] \ar[ruuuu]   & (d*F_5)_{t x z y_2 y_4 y_5} \\
    (C_4)_{y z y_2 y_3}  \ar[r] \ar[ruuuu] \ar[rd]  &   (F_5)_{u y z y_2 y_3}\ar[r]^{*} &   (*F_5)_{t x y_1 y_4 y_5} \ar[r] \ar[ruu] \ar[ruuuuu]   & (d*F_5)_{t x z y_1 y_4 y_5} \\
    (C_4)_{u y y_1 y_3}   \ar[ruu] \ar[ruuuu] \ar[rdd]   & (F_5)_{y z y_1 y_2 y_3}\ar[r]^{*} &   (*F_5)_{t u x y_4 y_5} \ar[ruuu] \ar[ruuuu] \ar[rdd]  & (d*F_5)_{u x z y_1 y_4 y_5}\\
(C_4)_{u y y_2 y_3} \ar[ruuuu] \ar[ruu] \ar[rd]   &   (F_5)_{t y y_1 y_2 y_3}\ar[r]^{*} &   (*F_5)_{ u x z y_4 y_5} \ar[ruuuuuu] \ar[rd] \ar[ru]   & (d*F_5)_{x z y_1 y_2 y_4 y_5} \\
   (C_4)_{y y_1 y_2 y_3}  \ar[r] \ar[ru] \ar[ruu]  & (F_5)_{u y y_1 y_2 y_3}\ar[r]^{*} &   (*F_5)_{t x z y_4y_5}  \ar[r] \ar[ruuu] \ar[ruuuu]   & (d*F_5)_{ t u x z y_4 y_5}\\
  }
\end{xy}
 \end{equation*}
\caption{Depiction of the system of differential equations, dual to those of diagram \ref{diag1}. Contributions of off-diagonal elements of the metric tensor to the Hodge duals were left out for simplicity in this figure, of course, they are included in the calculation. The right hand side of the diagram has to be equal to the corresponding directions of $d \big(* \frac{2 \gamma}{\sqrt{-g}}\frac{\delta \mathcal{W}}{\delta F_5}\big)$.}
\label{diag2}
 \end{figure}
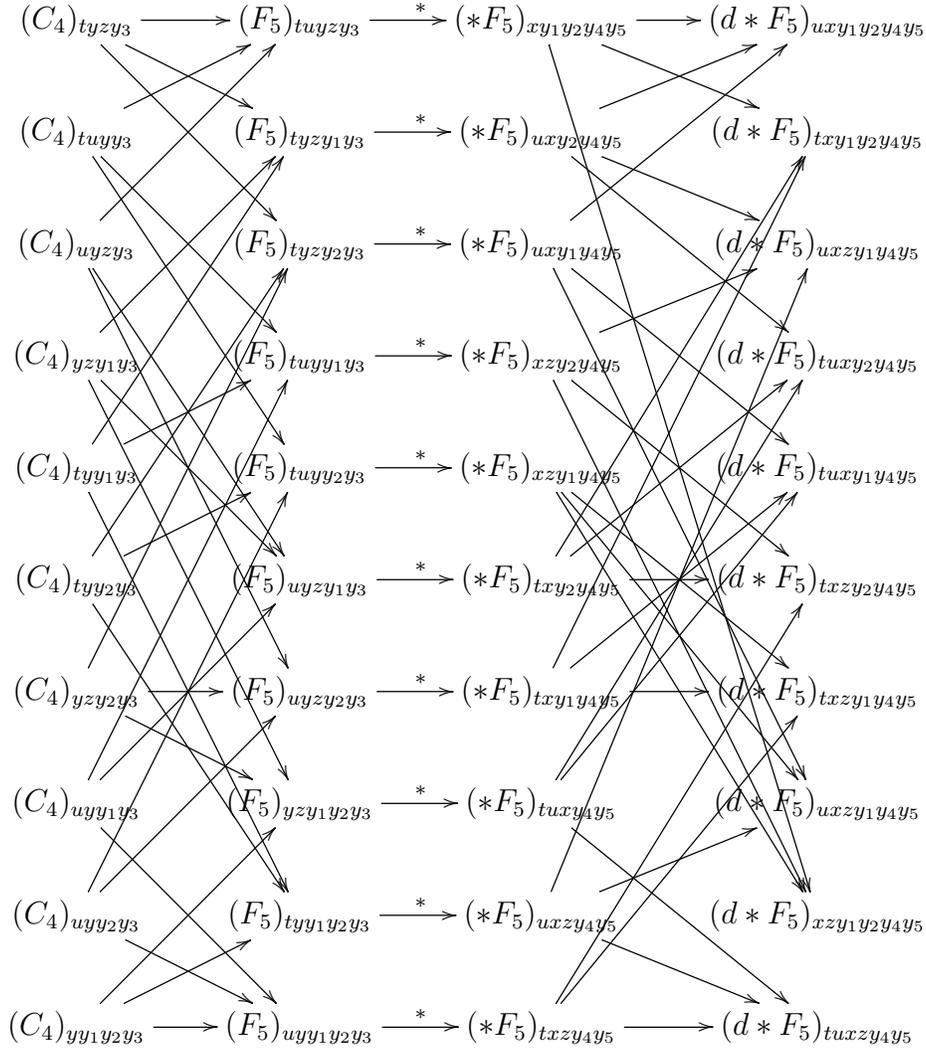

Now let us start to consider higher derivative corrections to our theory. In type IIb SUGRA this means that we have to add terms of order $\alpha'^3$ to the action (\ref{eq:action}). For this purpose we set $\gamma =\frac{\zeta(3)}{8}\lambda^{-\frac{3}{2}}$, with the 't Hooft coupling $\lambda$, which is proportional to $\alpha'^{-\frac{1}{2}}$. The action including finite $\lambda$ corrections has the form
\begin{equation}
S=S_{10}+\gamma S^{\gamma}_{10}+\mathcal{O}(\gamma^{\frac{4}{3}}),
\label{eq:action2}
\end{equation} 
with 
\begin{equation}
S_{10} = \frac{1}{2 \kappa_{10}} \int d^{10}x \sqrt{-g} \bigg[R_{10}-\frac{1}{4 \times 5!} F_5^2 \bigg].
\label{eq:action}
\end{equation}
as before and 
\begin{equation}
S^\gamma_{10}= \frac{1}{2\kappa_{10}} \int d^{10}x \sqrt{\vert g_{10}\vert} \bigg[C^4+C^3 \mathcal{T}+C^2  \mathcal{T}^2 +C \mathcal{T}^3+\mathcal{T}^4\bigg].
\label{actioncor}
\end{equation}
The expression for   $S^\gamma_{10}$ is   schematical and stands for a set of tensor contractions between the Weyl tensor $C$ and $\mathcal{T}$, a $6$-tensor that takes care of higher derivative corrections containing the five form. Explicitly the term in brackets in (\ref{actioncor}) is given by \cite{Paulos:2008tn}
\begin{equation}
\gamma W = \gamma\bigg[C^4+C^3 \mathcal{T}+C^2  \mathcal{T}^2 +C \mathcal{T}^3+\mathcal{T}^4\bigg]
 = \frac{\gamma}{86016} \sum_{i=1}^{20} n_i M_i,
\end{equation}
with
\begin{align}
\left(n_i\right)_{i=1,\dots, 20}=&(-43008,86016,129024,30240,7392,-4032,-4032,
-118272,\nonumber \\ &-26880,112896,-96768,1344,-12096,-48384,24192,2386,\nonumber \\ &-3669,
-1296,10368,2688 )
\end{align}
as well as
\begin{align}
(M_i)_{i=1,\dots,20}= &(C_{abcd} C^{abef}C^{c} \hspace{0.01cm}_{egh}C^{dg}\hspace{0.01cm}_f\hspace{0.01cm}^h, C_{abcd}C^{aecf} C^{bg}\hspace{0.01cm}_{eh}C^d\hspace{0.01cm}_{gf}
\hspace{0.01cm}^h,\nonumber \\ \nonumber & C_{abcd}C^a\hspace{0.01cm}_e\hspace{0.01cm}^f
\hspace{0.01cm}_g C^b\hspace{0.01cm}_{fhi}\mathcal{T}^{cdeghi},C_{abc}
\hspace{0.01cm}^d C^{abc}\hspace{0.01cm}_e\mathcal{T}_{dfghij}
\mathcal{T}^{efhgij},\nonumber \\ &
C_a\hspace{0.01cm}^{bcd}C^a\hspace{0.01cm}_{bef}
\mathcal{T}_{cdghij}\mathcal{T}^{efghij},
C_a\hspace{0.01cm}^{bc}\hspace{0.01cm}_dC^{ae}\hspace{0.01cm}_{cf}
\mathcal{T}_{beghij}\mathcal{T}^{dfghij}\nonumber  \\ &
C_a\hspace{0.01cm}^{bcd} C^a\hspace{0.01cm}_{ecf}\mathcal{T}_{bghdij}
\mathcal{T}^{eghfij},
C_a\hspace{0.01cm}^{bc}\hspace{0.01cm}_{d}
C^{ae}\hspace{0.01cm}_{fg}\mathcal{T}_{bcehij}
\mathcal{T}^{dfhgij}, \nonumber \\ &C_a\hspace{0.01cm}^{bc}\hspace{0.01cm}_{d}
C^{ae}\hspace{0.01cm}_{fg}\mathcal{T}_{bcehij}
\mathcal{T}^{dhifgj},C_a\hspace{0.01cm}^{bc}\hspace{0.01cm}_{d}
C^{a}\hspace{0.01cm}_e\hspace{0.01cm}^{f}\hspace{0.01cm}_g
\mathcal{T}_{bcfhij}
\mathcal{T}^{dehgij}, \nonumber \\ &C_a\hspace{0.01cm}^{bc}\hspace{0.01cm}_{d}
C^{ae}\hspace{0.01cm}_{fg}\mathcal{T}_{bcheij}
\mathcal{T}^{dfhgij}, C^{a b c d} \mathcal{T}_{abefgh}\mathcal{T}_{cd}\hspace{0.01cm}^{eijk}
\mathcal{T}^{fgh}\hspace{0.01cm}_{ijk},\nonumber \\ & C^{a b c d} \mathcal{T}_{abefgh}\mathcal{T}_{cd}\hspace{0.01cm}^{fijk}
\mathcal{T}^{egh}\hspace{0.01cm}_{ijk}, C^{a b c d} \mathcal{T}_{abefgh}\mathcal{T}_{cd}\hspace{0.01cm}^{fijk}
\mathcal{T}^{eg}\hspace{0.01cm}_i\hspace{0.01cm}^h
\hspace{0.01cm}_{jk}, \nonumber  \\ & C^{a b c d} \mathcal{T}_{abefgh}\mathcal{T}_{c}\hspace{0.01cm}^{efijk}
\mathcal{T}_{d}\hspace{0.01cm}^{gh}
\hspace{0.01cm}_{ijk},\mathcal{T}_{abcdef}\mathcal{T}^{abcdgh}\mathcal{T}^e
\hspace{0.01cm}_{gijkl}\mathcal{T}^{fij}\hspace{0.01cm}_h
\hspace{0.01cm}^{kl}, \nonumber  \\ &\mathcal{T}_{abcdef}\mathcal{T}^{abcghi}\mathcal{T}^{de}
\hspace{0.01cm}_{jg}\hspace{0.01cm}^{kl}\mathcal{T}^{f}
\hspace{0.01cm}_{hki}
\hspace{0.01cm}^{j}\hspace{0.01cm}_l,\mathcal{T}_{abcdef}\mathcal{T}^{abcghi}\mathcal{T}^{d}
\hspace{0.01cm}_{gj}\hspace{0.01cm}^{ekl}\mathcal{T}^{f}
\hspace{0.01cm}_{h}
\hspace{0.01cm}^{j}\hspace{0.01cm}_{ikl}\nonumber \\ &
\mathcal{T}_{abcdef}\mathcal{T}^{abcghi}\mathcal{T}^{d}
\hspace{0.01cm}_{gj}\hspace{0.01cm}^{ekl}\mathcal{T}^{f}
\hspace{0.01cm}_{hki}
\hspace{0.01cm}^{j}\hspace{0.01cm}_l, \mathcal{T}_{abcdef}\mathcal{T}^{aghdij}\mathcal{T}^{b}
\hspace{0.01cm}_{gk}\hspace{0.01cm}^e\hspace{0.01cm}_{il}\mathcal{T}^{c}\hspace{0.01cm}_{h}
\hspace{0.01cm}^{kf}\hspace{0.01cm}_{j}\hspace{0.01cm}^l
).
\end{align}
 The Weyl tensor $C_{abcd}$ is 
\begin{align}
C_{abcd} & \nonumber =R_{abcd}-\frac{1}{8} \big(g_{ac}R_{db}-g_{ad}R_{cb}-g_{bc} R_{da}+g_{bd}R_{ca} \big)+\frac{1}{72} \times \\ & \times \big(R g_{ac} g_{db}-R g_{ad}g_{cb} \big),
\end{align} and $\mathcal{T}$ is given by
\begin{equation}
\mathcal{T}_{abcdef}= i \nabla_a F_{bcdef}^++\frac{1}{16} \big(F_{abcmn}^+ F_{def}^{+ mn}-3 F_{abfmn}^+ F_{dec}^{+ mn}\big),
\label{Tt}
\end{equation}
with two sets of antisymmetrized indices $a,b,c$ and $d,e,f$. In addition the right hand side of (\ref{Tt}) is  symmetrized with respect to the interchange of $(a,b,c) \leftrightarrow (d,e,f)$    ~\cite{Paulos:2008tn}.  Here $F^+$ stands for the self dual part $\frac{1}{2}(1+*)F_5$ of the five form. It should be noted that up to this day it is not known, whether the terms in (\ref{actioncor}), which were derived in \cite{Paulos:2008tn} using \cite{Skenderis}, are complete. There are strong indications that this is the case, but since there is no strict mathematical proof we included this cautionary remark. \\ \indent We already know that the solution of $F_5$ in order $\mathcal{O}(\gamma^0)$ is self dual, and that in order $\mathcal{O}(\gamma^1)$ the $\mathcal{O}(\gamma^0)$ part of  $F_5$ is the only contribution of $F_5$ that enters in the higher derivative part of the action. But we still do not have the EoMs in order $\mathcal{O}(\gamma)$  for the 4-form components. This means that we still have to vary the action with respect to $C_4$ and thus it makes a difference whether   $F_5= dC_4$ or $F^+$ enters $\gamma W$. Before we start discussing the higher derivative corrected EoMs for gauge fields, we have to determine the $\gamma$-corrected solution of our unperturbed geometry as done in  \cite{Pawelczyk:1998pb}. The  ansatz for the metric we make is of the form 
\begin{align}
ds_{10}^2= &-r_h^2 U(u)dt^2+ \tilde{U}(u) du^2+ e^{2V(u)}r_h^2(dx^2+dy^2+dz^2)+L(u)^2 d\Omega_5^2,
\label{metric}
\end{align}
where we are forced to give up the product structure of our manifold and admit a $u$-dependent warping factor $L(u)$ in front of the $5$-sphere line element as shown in \cite{Pawelczyk:1998pb}. The EoMs for our $4$-form components still have the form (\ref{eqF51}) simply because the $\mathcal{T}$-tensor defined above vanishes on the unperturbed background.  We also have 
\begin{equation}
\frac{\delta S_{10}^\gamma}{\delta F_5}=0
\end{equation}
for $A_\mu =0$. The solution for the $5$-form in order $\mathcal{O}(\gamma^1)$ and without gauge fields is 
\begin{equation}
F_5=(1+*)F_5^{el}
\end{equation}
\begin{equation}
F_5^{el}=\frac{-4}{L(u)^5}\epsilon_{AdS}^{\gamma},
\end{equation}
where $\epsilon_{AdS}^{\gamma}$ is the volume form of the $\gamma$-corrected AdS-part of our manifold. The five form is still self dual, such that we are allowed to plug the solution for the five form back into the action, only considering its magnetic part and doubling its contribution, which gives 
 \begin{equation}
\frac{1}{2\kappa_{10}}\int d^{10}x\sqrt{-\det(g_{10})}\bigg[R_{10} -\frac{8}{L(u)^{10}}+\gamma W\bigg].
\end{equation}
The EoM for the metric components from this action yield \cite{Pawelczyk:1998pb}
\begin{align}
&U(u)=\frac{(1-u^2)}{u}\bigg(1+\frac{5 u^2 \gamma}{8} (-130 - 130 u^2 + 67 u^4) \bigg)\\ &
\tilde{U}(u)=\frac{1}{4 u^2 (1-u^2)}\bigg(1+\gamma\Big(\frac{325}{4} u^2 + \frac{1075}{16} u^4 -\frac{ 4835}{16} u^6 \Big)\bigg)
\\& V(u)=-\frac{1}{2}\log(u)\\&
L(u)=1+\frac{15\gamma }{32}(1+u^2)u^4.
\end{align}
Now we are ready to introduce gauge fields to our finite $\lambda$-corrected theory. In order to get the correct results in the limits $A_\mu \to 0$ and $\gamma \to 0$ we choose the ansatz again corresponding to a twist of the five sphere along the $y_3$, $y_4$, $y_5$ angles
\begin{align}
ds_{10}^2= &-r_h^2 U(u)dt^2+ \tilde{U}(u) du^2+ e^{2V(u)}r_h^2(dx^2+dy^2+dz^2)+L(u)^2 \frac{4 A_x(u,t,z)^2}{3} dx^2\nonumber \\&+L(u)^2 \frac{4A_x(u,t,z)}{\sqrt{3}}dx \big(dy_3 \sin(y_1)^2 +dy_4 \cos(y_1)^2\sin(y_2)^2+dy_5 \cos(y_1)^2\nonumber \\ &\cos(y_2)^2  \big)+
L(u)^2\big(dy_1^2+\cos(y_1)^2 dy_2^2+\sin(y_1)^2dy_3^2+\cos(y_1)^2\sin(y_2)^2 dy_4 \nonumber \\& +\cos(y_1)^2\cos(y_2)^2 dy_5^2\big),
\label{eq:metricc}
\end{align}
which we  justify as follows: We will obtain the EoM for $A_\mu$ by varying the coupling corrected type IIb SUGRA action with respect to the $4$-form components and $A_\mu$. Apparently the $xx$-component $e^{2V(u)}+L(u)^2 \frac{4 A_x(u,t,z)^2}{3}$ of our metric ansatz  looks like it could lead to problems. On the one hand we know that if we would only vary the action with respect to e.g. the $x y_3$-component, we would  obtain an EoM for $A_\mu$ that is at first glance different from varying the action with respect to $A_\mu$. This is because after linearizing in $A_\mu$ the $A_\mu^2$-term of the $xx$-component of the metric won't contribute to the former case, but will give a contribution to the latter. In fact varying the $\sqrt{-g}R_{10}$, $\sqrt{-g} \frac{1}{4*5!}F_5^2$ and $\sqrt{-g}\gamma W$-terms in the action with respect to the  $xy_3$-component of the metric separately and inserting the ansatz (\ref{eq:metricc}) gives mass terms. However, adding everything up leads to the same EoM for $A_\mu$ (of course, still depending on some unknown $F_5$-directions) as varying with respect to $A_\mu$, while the mass terms cancel identically. \\ \indent 
 From now on we will work with $r_h=1$, which also applies for the Appendix \ref{Appendix}, and reintroduce $r_h$ wherever needed after having obtained the EoM or contributions thereto. We know that we will end up with  differential equations, where $r_h$ only appears in the rescaled frequency $\frac{\omega}{2 r_h}$ and momentum $\frac{q}{ 2r_h}$. Also setting $r_h=1$ simply corresponds to rescaling $t$ the spatial coordinates and $A_x$ by a constant factor. Changing $\frac{\omega}{2}$ to $\frac{\omega}{2 r_h}$ in the end corresponds to scaling back to the form of the metric given in  (\ref{eq:metricc}). \\ \indent 
Now we are prepared to determine the EoM in order $\mathcal{O}(\gamma)$  of all relevant fields i.e. gauge fields, the five-form and, less important, the dilaton field. Since its EoM decouple, we will ignore it henceforth. Let us start with the five-form. As in the last section its EoMs are derived by varying the action with respect to the $4$-form components with $d C_4=F_5$. A concise way of writing the resulting system of differential equations is
\begin{equation}
d \bigg(*F_5-*\frac{2\gamma}{\sqrt{-g}}\frac{\delta \mathcal{W}}{\delta F_5} \bigg)=0,
\label{EoMF5}
\end{equation}
where $\frac{\delta \mathcal{W}}{\delta F_5} $ is defined by
\begin{equation}
\frac{\delta \mathcal{W}}{\delta F_5}:=2 \kappa_{10}\frac{\delta S_{10}^\gamma}{\delta F_5}.
\end{equation}
It is easy to obtain this by observing that for a $p$-form $C$ with $F=dC$ and an action
\begin{equation}
S=\int d^Dx L(F, \nabla F)
\end{equation}
for $C$ the variation
$
\frac{\delta S}{\delta C}=0
$
leads to an equivalent set of differential equations as 
\begin{equation}
d \bigg( *\frac{1}{\sqrt{-g}} \frac{\delta S}{\delta F}\bigg)=0.
\end{equation}
The first and easiest result we can extract from (\ref{EoMF5}) is that self duality of the five form is broken if $d*\frac{1}{\sqrt{-g}}\frac{\delta \mathcal{W}}{\delta F_5}\neq 0$, which is the case if $A_\mu \neq 0$. Obviously, if $F_5$ would still be self dual, we had $(1-*)F_5=0$, but together with $dF_5=0$ (\ref{EoMF5}) would then lead to a contradiction. This means that  we cannot treat the $F_5^2$-term of the action as in the previous cases. In the following let us focus on the variation of this term with respect to $A_\mu$. \\ \indent Due to the same argument as in the first section, since we are only interested in those terms of the final EoM, which are linear in $A_\mu$, we can ignore $\mathcal{O}(A_\mu ^2)$ parts of the metric in $F_5^2$. Contributions of terms of this form cancel  identically, as they  have to, since otherwise we would get mass terms. This means that the number of $F_5$-directions, which actually contribute to 
\begin{equation}
 \frac{\delta \sqrt{-g}F_5^2}{\delta A_\mu}
 \label{var1}
\end{equation}
is very restricted. As in section one, we only consider transverse fields $A_x(u,t,z)$, with $A_y=A_z=0$. This implies that the only metric components depending on $A_\mu$, modulo terms of order $\mathcal{O}(A_\mu^2)$, are again $g^{x y_3}, g^{x y_4}, g^{x y_5}, g^{y_3 x}, g^{y_4 x}, g^{y_5 x}$. Therefore, the only directions of $F_5$, which contribute to (\ref{var1}) in order $\mathcal{O}(\gamma^1)$, are 
\begin{align}
& (F_5)_{y_1 y_2 y_3 y_4 y_5},(F_5)_{ tuxyz},(F_5)_{tuyz y_3},(F_5)_{tuyz y_4},(F_5)_{tuyz y_5},(F_5)_{x y_1 y_2  y_4 y_5},(F_5)_{x y_1 y_2  y_3 y_5},\nonumber  \\& (F_5)_{x y_1 y_2  y_3 y_4}.
\label{directions}
\end{align}
We already know how $(F_5)_{y_1 y_2 y_3 y_4 y_5}$ and $(F_5)_{ tuxyz}$ look like in order $\mathcal{O}(\gamma^1)$ for $A_\mu=0$ and  how these directions are modified  in order $\mathcal{O}(\gamma^0)$ for $A_\mu \neq0$. This is  all the information we need about them, when computing (\ref{var1}), since $(F_5)_{tuyz y_3},(F_5)_{tuyz y_4}$, $(F_5)_{tuyz y_5}$, $(F_5)_{x y_1 y_2  y_4 y_5}$, $(F_5)_{x y_1 y_2  y_3 y_5}$, $ (F_5)_{x y_1 y_2  y_3 y_4}$ are zero for $A_\mu=0$. This means we only  have to compute $(F_5)_{tuyz y_3}$, $(F_5)_{tuyz y_4}$, $(F_5)_{tuyz y_5}$, $(F_5)_{x y_1 y_2  y_4 y_5}$, $(F_5)_{x y_1 y_2  y_3 y_5}$, $ (F_5)_{x y_1 y_2  y_3 y_4}$  up to first order in $\gamma$  from (\ref{EoMF5}). We will return to this later, let us first finish the variation of the rest of the action with respect to the gauge fields.\\ \indent 
With our metric (\ref{eq:metricc}) we obtain
\begin{equation}
R_{10}= \bigg(R_{10}\big|_{A_\mu \to 0}\bigg)-\frac{ L(u)^2}{3} F_{\mu \nu}F^{\mu \nu}
\end{equation}
for the Ricci scalar. Varying this part with respect to $A_\mu$ is straightforward. The final part
\begin{equation}
\frac{\delta \gamma \sqrt{-g} \mathcal{W}}{\delta A_\mu }
\end{equation} already  contains a $\gamma$-factor. Therefore, only $\mathcal{O}(\gamma^0)$-parts of the metric and $F_5$ enter it in order $\mathcal{O}(\gamma^1)$. Knowing already the solutions for $F_5$ with gauge fields in zeroth order in $\gamma$ allows us to compute this term immediately. One has to be careful and remember that only the self dual part of $F_5$ enters here. Of course, we know already, that after having solved all EoM, we have $(1-*)F_5=0$ in order $\mathcal{O}(\gamma^0)$. But since on the action level the $4$-form components and the gauge fields are independent fields, meaning that $\frac{\delta F_5}{\delta A_\mu}=0$, it is crucial to realize, that in  general
\begin{equation}
\frac{\delta f(\frac{1}{2}(1+*)F_5)}{\delta A_\mu} \neq\frac{\delta f(F_5)}{\delta A_\mu} 
\end{equation}
for a functional $f$, even if $\frac{1}{2}(1+*)F_5=F_5$ after inserting all solutions of the resulting EoM. This is because  $A_\mu$ can enter through to the Hodge dual
\begin{equation}
\frac{\delta (*F_5)_{a b c d e f}}{\delta A_\mu} \neq 0 =\frac{\delta (F_5)_{a b c d e f}}{\delta A_\mu} 
\end{equation}
for some directions $a b c de f$. Let us split the work up and concentrate on the $C^4$-part of the higher derivative corrections first. After varying it with respect to $A_x$, introducing 
\begin{equation}
(A_x)_k(u,q,\omega)=\frac{1}{2 \pi}\int dt dz e^{iq z}e^{-iw t}A_x(u,z,t)
\end{equation}
and exploiting that 
\begin{equation}
\gamma (\partial_u^2 A_x-\frac{2u}{1-u^2}\partial_u A_x+\frac{\hat{\omega}^2-\hat{q}^2(1-u^2)}{u (1-u^2)^2}A_x)=\mathcal{O}(\gamma^2)
\end{equation}
we obtain 
\begin{align}
&\frac{64u^3 \gamma }{3}  \big((A_x)_k (24 \hat q^4 u+\hat q^2 (162-235 u^2)-60
   \hat w^2)-(u^2-1) (120 \hat q^2 u-135 u^2+ \nonumber \\ & 112)
  (A_x)_k'\big)+\mathcal{O}(\gamma^2)
  \label{C4}
\end{align}
as a contribution to the differential equations, rescaled in such a way that the $\mathcal{O}(\gamma^0)$-part has the form $\frac{8(1-u^2)}{3}$ times (\ref{eq0}). We can ignore terms of the form $C \mathcal{T}^3$, $\mathcal{T}^4$, since we are only interested in linearized EoMs for $A_x$ and $\mathcal{T}=0$ on a fluctuation free metric. However, since $\frac{\delta \mathcal{T}}{\delta A_\mu}\neq 0$, we still have to determine \begin{equation}
\frac{\delta \gamma \sqrt{-g} C^2 \mathcal{T}^2}{\delta A_x}\hspace{0.3 cm} \text{and} \hspace{0.3 cm} \frac{\delta \gamma \sqrt{-g} C^3 \mathcal{T}}{\delta A_x} .
\label{var2}
\end{equation} Our strategy to compute the terms above will be to insert
the solutions for $F_5$ in lowest order in $\gamma$ slightly modified by replacing $A_\mu$ by a new independent function $\bar A_\mu$ into (\ref{var2})
\begin{equation}
F_5\bigg|_{A_\mu \to \bar{A}_\mu}
\end{equation} 
and let $\bar{A}_\mu$ go to $A_\mu$ after the variation, since we are not allowed to vary with respect to $A_\mu$ appearing in $F_5$ after inserting the $\mathcal{O}(\gamma^0)$ solution of the five form. For this purpose let us write down explicitly how this solution  looks like.
We start with the gauge field free electric and magnetic part and get
\begin{align}
(F_5^{el})^0&= -4\sqrt{|\det(g_5)|} dt\wedge du \wedge dx\wedge dy \wedge dz\nonumber \\ 
*(F_5^{el})^0&= 4 \sqrt{\det(g_{S_5})}dy_1\wedge dy_2 \wedge  dy_3 \wedge dy_4 \wedge dy_5+4 \sqrt{|\det(g_{10})|}\sqrt{|\det(g_5)|}\nonumber \\ & \bigg(g_{10}^{t t}g_{10}^{u u}g_{10}^{yy}g_{10}^{x y_3}g_{10}^{z  z} dy_1 \wedge dy_2 \wedge dx \wedge dy_4 \wedge dy_5 +g_{10}^{t t}g_{10}^{u u}g_{10}^{yy}g_{10}^{x y_4}g_{10}^{zz } dy_1 \wedge dy_2   \nonumber \\ &\wedge dy_3\wedge dx \wedge dy_5+g_{10}^{t  t}g_{10}^{u  u}g_{10}^{ y y }g_{10}^{x y_{5}}g_{10}^{ z z} dy_1 \wedge dy_2 \wedge dy_3 \wedge dy_4 \wedge dx\bigg)=:(F_5^{mag})^0,
\end{align}
where $g_{10}$ is the metric of the $10$ dimensional manifold corresponding to an AdS-Schwarzschild black hole times $S_5$, $g_5$ is the metric corresponding to the internal AdS space and $g_{S_5}$ is the metric of the five sphere. The nomenclature $(F_5^{mag})^0$ shouldn't distract from the fact that it nevertheless depends on $A_\mu$  via $g_{10}^{x y_{5}}$, $g_{10}^{x y_{4}}$ and
$g_{10}^{x y_{3}}$. The electric components of the five form including the gauge field $A_x(u,t,z)$ are explicitly given by 
\begin{equation}
(F_5^{el})^1=(F_5^{el})^1_{ux}+(F_5^{el})^1_{tx}+(F_5^{el})^1_{zx}
\end{equation}
with
\begin{align}
(F_5^{el})^1_{ux}&= \frac{2 \partial_u A_x(u,t,z)}{ \sqrt{3}}\sqrt{|\det(g_5)|}g_5^{xx} g_5^{uu }\big(\sin(y_1)\cos(y_1)dt \wedge dy \wedge dz \wedge dy_1 \wedge dy_3 +\nonumber \\ & \cos(y_1)^2\sin(y_2)\cos(y_2)dt \wedge dy \wedge dz \wedge dy_2 \wedge dy_4-\cos(y_1)\sin(y_1)\sin(y_2)^2dt \nonumber \\ &\wedge dy \wedge dz \wedge dy_1 \wedge dy_4-\cos(y_1)\sin(y_1)\cos(y_2)^2dt \wedge dy \wedge dz \wedge dy_1 \wedge dy_5\nonumber \\ &-\cos(y_2)\sin(y_2)\cos(y_1)^2dt \wedge dy \wedge dz \wedge dy_2 \wedge dy_5
 \big),
 \label{eq:Fhodgestart}
\end{align}
\begin{align}
(F_5^{el})^1_{tx}&= -\frac{2 \partial_t A_x(u,t,z)}{ \sqrt{3}}\sqrt{|\det(g_5)|}g_5^{xx} g_5^{tt }\big(\sin(y_1)\cos(y_1)du \wedge dy \wedge dz \wedge dy_1 \wedge dy_3 +\nonumber \\ & \cos(y_1)^2\sin(y_2)\cos(y_2)du \wedge dy \wedge dz \wedge dy_2 \wedge dy_4-\cos(y_1)\sin(y_1)\sin(y_2)^2du \nonumber \\ &\wedge dy \wedge dz \wedge dy_1 \wedge dy_4-\cos(y_1)\sin(y_1)\cos(y_2)^2du \wedge dy \wedge dz \wedge dy_1 \wedge dy_5\nonumber \\ &-\cos(y_2)\sin(y_2)\cos(y_1)^2du \wedge dy \wedge dz \wedge dy_2 \wedge dy_5
 \big),
\end{align}
\begin{align}
(F_5^{el})^1_{zx}&=- \frac{2 \partial_z A_x(u,t,z)}{ \sqrt{3}}\sqrt{|\det(g_5)|}g_5^{xx} g_5^{zz }\big(\sin(y_1)\cos(y_1)dt \wedge dy \wedge du \wedge dy_1 \wedge dy_3 +\nonumber \\ & \cos(y_1)^2\sin(y_2)\cos(y_2)dt \wedge dy \wedge du \wedge dy_2 \wedge dy_4-\cos(y_1)\sin(y_1)\sin(y_2)^2dt \nonumber \\ &\wedge dy \wedge du \wedge dy_1 \wedge dy_4-\cos(y_1)\sin(y_1)\cos(y_2)^2dt \wedge dy \wedge du \wedge dy_1 \wedge dy_5\nonumber \\ &-\cos(y_2)\sin(y_2)\cos(y_1)^2dt \wedge dy \wedge du \wedge dy_2 \wedge dy_5
 \big).
\end{align}
Analogously we write the magnetic part as
\begin{equation}
(F_5^{mag})^1=(F_5^{mag})^1_{ux}+(F_5^{mag})^1_{tx}+(F_5^{mag})^1_{zx}
\end{equation}
with
\begin{align}
(F_5^{mag})^1_{ux}&= -\sqrt{\det(g_{S_5})}\big(\sin(y_1)\cos(y_1)g_{10}^{y_1 y_1}g_{10}^{y_3 y_3} du \wedge dx \wedge dy_2 \wedge dy_5 \wedge dy_4 +\nonumber \\ &  \cos(y_1)^2\sin(y_2)\cos(y_2)g_{10}^{y_2 y_2}g_{10}^{y_4 y_4} du \wedge dx \wedge dy_1 \wedge dy_5 \wedge dy_3-\sin(y_1)\times \nonumber \\ &\cos(y_1)\sin(y_2)^2g_{10}^{y_1 y_1}g_{10}^{y_4 y_4} du \wedge dx \wedge dy_2 \wedge dy_3 \wedge dy_5-\cos(y_1) \sin(y_1)\times \nonumber \\ &\cos(y_2)^2g_{10}^{y_1 y_1}g_{10}^{y_5 y_5} du \wedge dx \wedge dy_2 \wedge dy_4 \wedge dy_3-\cos(y_2) \sin(y_2)\cos(y_1)^2\times \nonumber  \\ &g_{10}^{y_2 y_2}g_{10}^{y_5 y_5} du \wedge dx \wedge dy_1 \wedge dy_3 \wedge dy_4)\frac{2 \partial_u A_ x(u,t,z)}{\sqrt 3}+\mathcal{O}(A_x(u,t,z)^2),
\end{align}
\begin{align}
(F_5^{mag})^1_{tx}&= -\sqrt{\det(g_{S_5})}\big(\sin(y_1)\cos(y_1)g_{10}^{y_1 y_1}g_{10}^{y_3 y_3} dt \wedge dx \wedge dy_2 \wedge dy_5 \wedge dy_4 +\nonumber \\ &  \cos(y_1)^2\sin(y_2)\cos(y_2)g_{10}^{y_2 y_2}g_{10}^{y_4 y_4} dt \wedge dx \wedge dy_1 \wedge dy_5 \wedge dy_3-\sin(y_1)\times \nonumber \\ &\cos(y_1)\sin(y_2)^2g_{10}^{y_1 y_1}g_{10}^{y_4 y_4} dt \wedge dx \wedge dy_2 \wedge dy_3 \wedge dy_5-\cos(y_1) \sin(y_1)\times \nonumber \\ &\cos(y_2)^2g_{10}^{y_1 y_1}g_{10}^{y_5 y_5} dt\wedge dx \wedge dy_2 \wedge dy_4 \wedge dy_3-\cos(y_2) \sin(y_2)\cos(y_1)^2\times \nonumber  \\ &g_{10}^{y_2 y_2}g_{10}^{y_5 y_5} dt \wedge dx \wedge dy_1 \wedge dy_3 \wedge dy_4)\frac{2 \partial_t A_ x(u,t,z)}{\sqrt 3}+\mathcal{O}(A_x(u,t,z)^2),
\end{align}
\begin{align}
(F_5^{mag})^1_{zx}&= -\sqrt{\det(g_{S_5})}\big(\sin(y_1)\cos(y_1)g_{10}^{y_1 y_1}g_{10}^{y_3 y_3} dz \wedge dx \wedge dy_2 \wedge dy_5 \wedge dy_4 +\nonumber \\ &  \cos(y_1)^2\sin(y_2)\cos(y_2)g_{10}^{y_2 y_2}g_{10}^{y_4 y_4} dz \wedge dx \wedge dy_1 \wedge dy_5 \wedge dy_3-\sin(y_1)\times \nonumber \\ &\cos(y_1)\sin(y_2)^2g_{10}^{y_1 y_1}g_{10}^{y_4 y_4} dz \wedge dx \wedge dy_2 \wedge dy_3 \wedge dy_5-\cos(y_1) \sin(y_1)\times \nonumber \\ &\cos(y_2)^2g_{10}^{y_1 y_1}g_{10}^{y_5 y_5} dz\wedge dx \wedge dy_2 \wedge dy_4 \wedge dy_3-\cos(y_2) \sin(y_2)\cos(y_1)^2\times \nonumber  \\ &g_{10}^{y_2 y_2}g_{10}^{y_5 y_5} dz \wedge dx \wedge dy_1 \wedge dy_3 \wedge dy_4)\frac{2 \partial_z A_ x(u,t,z)}{\sqrt 3}+\mathcal{O}(A_x(u,t,z)^2).
 \label{eq:Fhodgeend}
\end{align}
The complete solution of the five form $F_5$ in order $\mathcal{O}(\gamma)$ is then
\begin{equation}
F_5=(F_5^{mag})^1+(F_5^{mag})^0+(F_5^{el})^1+(F_5^{el})^0.
\end{equation}
One easy way of testing this five form solution is to compute $F_5^2$, which turns out to be zero. This is good news, since the  Hodge star operator fulfills  for any five form $F$:
\begin{equation}
F\wedge *F=F^2 \tilde \omega,
\end{equation} 
where $\tilde \omega$ is the $10$ form
\begin{equation}
\tilde \omega =dt \wedge du \wedge dx \wedge dy \wedge dz \wedge dy_1 \wedge dy_2 \wedge dy_3
\wedge dy_4
\wedge dy_5.
\end{equation}
Since $F_5$ is self dual in this order in $\alpha'$, we thus have to get $F_5^2=0$. It should be noted that this, of course, does not hold on the action level, even in the lowest order in $\alpha'$, since 
\begin{equation}
\bigg(\frac{\delta}{\delta A_\mu}\int d^{10} \sqrt{-g}( F_5|_{A_\mu \to \bar A_\mu})^2\bigg)_{\bar A_\mu \to A_\mu}
\end{equation}
does not have to vanish even if $F_5^2=0$ after inserting its solution.
Now let us think about which directions of $*F_5$ can actually enter (\ref{var2}). The only way $A_\mu$ can enter $C^3\mathcal{T}$ and $C^2 \mathcal{T}^2$ is through the fact that
\begin{equation}
\frac{\partial *( F_5|_{A_\mu \to \bar A_\mu})}{\partial A_\mu}\bigg|_{\bar{A}\mu \to A_\mu} \neq 0,
\label{dif1}
\end{equation} 
 $A_\mu$-dependent terms entering directly via the metric components present in the contractions of $C$ and $\mathcal{T}$, the Weyl tensor itself and the covariant derivative in (\ref{Tt}).  We claim that all we have to care about, regarding $*(F_5|_{A_\mu  \to \bar {A}_{\mu}})$ in (\ref{dif1}) is
\begin{equation}
F_5|_{A_\mu  \to \bar {A}_{\mu}}+ \frac{*(F_5^{el})^0-\big((F_5^{mag})^0|_{A_\mu  \to \bar {A}_{\mu}}\big)}{2}+\frac{*\big((F_5^{mag})^0|_{A_\mu  \to \bar {A}_{\mu}}\big)-\big(*(F_5^{mag})^0|_{A_\mu  \to \bar {A}_{\mu}}\big)}{2}.
\label{F5enter}
\end{equation} 
We also checked this explicitly by computing the unsimplified contribution of $*(F_5|_{A_\mu  \to \bar {A}_{\mu}})$ and explain in the following why this holds.\\ \indent
It is easy to see that this is true for 
the first term in (\ref{var2}). There, the argument that $\mathcal{T}=0$ for $\gamma=0$ and $A_\mu=0$  forces all contribution of order $\mathcal{O}(A_\mu^2)$, $\mathcal{O}(A_\mu  \bar A_\mu)$  or $\mathcal{O}(A_\mu\partial \bar A_\mu)$ from $* (F_5|_{A_\mu \to \bar A_\mu})$ to  $C^2 \mathcal{T}^2$  to be  negligible in (\ref{var2}). But what about potential terms of order $\mathcal{O}(A_\mu \partial \bar{A}_\mu)$ in $* (F_5|_{A_\mu \to \bar A_\mu})$ entering $C^3\mathcal{T}$? In fact, since the perturbation of the metric by $A_\mu$ was chosen in such a maximally symmetric way, in order to avoid coupling to scalars in order $\mathcal{O}(\gamma^0)$, it is rather straightforward to check that the terms of order $\mathcal{O}(A_\mu \partial \bar{A}_\mu)$ from $*((F_5^{mag})^1|_{A_\mu \to \bar{A}_{\mu}})$ cancel identically. Considering the definition of the tensor $\mathcal{T}$ one sees that the terms $*((F_5^{el})^1|_{A_\mu \to \bar{A}_{\mu}})$  of order $\mathcal{O}(A_\mu \partial \bar{A}_\mu)$ only enter those components $\mathcal{T} _{ a bc de f}$, where at least  one of $a, b,c, d,e$ is in $ \{y_1,\dots,y_5\}$. The parts of $\mathcal{T}$ coming from $*((F_5^{el})^1|_{A_\mu \to \bar{A}_{\mu}})$ in order $\mathcal{O}(A_\mu \partial \bar{A}_\mu)$, have to be contracted with the Weyl-tensor part of (\ref{var2}) computed from the $\gamma=0$-$A_\mu=0$-metric. In this case the Weyl tensor splits up block-diagonally into an AdS-part and a $S_5$-part, the latter of which is zero since the $5$-sphere is Weyl flat.  Summing up the contributions of both terms in (\ref{var2}) to the EoM obtained by variation with respect to $A_\mu$ one gets
\begin{equation}
\frac{16}{9} u^3 \left(349 \hat{q}^2 (A_x)_k-1111 \left(u^2-1\right) \partial_u (A_x)_k\right)+\mathcal{O}(\gamma^2).
\end{equation}
This term is rescaled in the same way  as (\ref{C4}). \\ \indent Now we have to solve (\ref{EoMF5}) for the last $6$ elements of (\ref{directions}). Again, as in the case $\gamma=0$, the strategy is to find closed diagrams such as figure \ref{diag1}, for which \begin{equation}
(F_5)_{tuyz y_3},(F_5)_{tuyz y_4},(F_5)_{tuyz y_5},(F_5)_{x y_1 y_2  y_4 y_5},(F_5)_{x y_1 y_2  y_3 y_5}, (F_5)_{x y_1 y_2  y_4 y_4}
\end{equation}
contribute to all considered directions of $d * F_5$  on the right side of the diagram and no more. After that one has to find all directions of $C_4$ that contribute to these components of $d * F_5$   and make sure that the directions of $C_4$ on the left side of the diagram don't contribute to another direction of $d * F_5$, otherwise expand the diagram and repeat. Let's assume we thereby collect a set of directions $\{a_i b_i c_i d_i e_i f_i\}_{i \in I}$ for which $(d * F_5)_{a_i b_i c_i d_i e_i f_i}$ with $i \in I$ appears on the right side of one of the diagrams. This means that we have to compute the components 
 \begin{equation}
 \bigg\{ \bigg(d * \frac{2 \gamma}{\sqrt{-g}}\frac{\delta \mathcal{W}}{\delta F_5}\bigg)_{a_i b_i c_i d_i e_i f_i}\bigg \}_{i \in I}
 \end{equation}
 in order to be able to solve for all needed directions of (\ref{EoMF5}). What we have to keep in mind is that in this order in $\gamma$ the five form is no longer self dual, such that we cannot simply skip one half of the diagrams and determine the remaining directions of $F_5$ with the help of the duality argument, as done in order $\mathcal{O}(\gamma^0)$. Since diagram \ref{diag1} was found without using that we are in order $\mathcal{O}(\gamma^0)$, we can simply reuse it now. But as said above we also have to find its dual diagram, which is given by figure \ref{diag2}. Here the unlabeled arrows in the diagram on the left and right depict derivatives. Due to (\ref{EoMF5}) the nonzero directions of $ d * \frac{2 \gamma}{\sqrt{-g}}\frac{\delta \mathcal{W}}{\delta F_5}$ determine the $y_1, y_2$-dependence of the components of $C_4$ proportional to $\gamma$ on the left hand side of the diagram.   The form of the solution of the five form in order $\mathcal{O}(\gamma^0)$, which gives the $y_1,y_2$-dependence of $ d * \frac{2 \gamma}{\sqrt{-g}}\frac{\delta \mathcal{W}}{\delta F_5}+ \mathcal{O}(\gamma^2)$ already illustrates, what becomes more apparent once one calculated the 
\begin{align*}
u x y_1 y_2 y_4 y_5,& t x y_1 y_2 y_4 y_5, u x z y_1 y_4y_5,t u x y_2 y_4y_5,t u x y_1 y_4 y_5,t x z y_2 y_4 y_5,\\ &t x z y_1 y_4 y_5,u x z y_1 y_4 y_5,x z y_1 y_2 y_4 y_5,t u x z y_4 y_5-
\end{align*}
directions of $ d * \frac{2 \gamma}{\sqrt{-g}}\frac{\delta \mathcal{W}}{\delta F_5}+ \mathcal{O}(\gamma^2)$, namely that all directions of $C_4$ on the left hand side, which contain a $y_2$ and all directions of $d*F_5$ on the right hand side, which contain a $y_1$ and no $y_2$ can be ignored, since all are trivially zero in order $\mathcal{O}(\gamma^1)$. More specifically we have \begin{equation}
 \bigg(d * \frac{2 \gamma}{\sqrt{-g}}\frac{\delta \mathcal{W}}{\delta F_5}\bigg)_{a bc y_1 y_4 y_5}= \mathcal{O}(\gamma^2)
\end{equation}
for all $a,b,c \in\{t,u,x,y,z,y_1,y_3,\dots,y_5\}$. 
 Before we turn to actually solving the differential equations linked to this diagram, its dual and  four further ones let us shortly address how to compute the differential form $\frac{\delta \mathcal{W}}{\delta F_5}$. To begin with,  one interesting observation is that
\begin{equation}
\frac{\delta \mathcal{W}}{\delta F_5}=\frac{1}{2}\bigg( 1-* \bigg)\frac{\delta \mathcal{W}}{\delta F_5^+},
\end{equation}
since only the self dual part of $F_5$ is entering $\gamma \mathcal{W}$. This relation could be used to test the result, once we have it, since it means that whatever we will obtain for $\frac{\delta \mathcal{W}}{\delta F_5}$ has to be anti-self dual.
In order to vary $\mathcal{W}$ or more specifically 
\begin{equation}
\int dx^{10} \sqrt{-g} \gamma W
\end{equation}
 we think of $W$  as a map
\begin{align}
W: \Omega^5(\mathcal{M}) \rightarrow C^\infty (\mathcal{M})
\end{align}
from the set of the $5$-forms on the manifold $\mathcal{M}$, which denotes the pseudo-riemannian manifold with metric (\ref{eq:metricc}), to $C^\infty (\mathcal{M})$. In order to compute the component $\frac{\delta \mathcal{W}}{\delta F_5}^{\mu_1,\dots,\mu_5}$ we take the limit
\begin{equation}
\lim_{\alpha \to 0} \frac{1}{\alpha} \int dx^{10} \sqrt{-g} \gamma \bigg(W[F_5+\alpha F(u,t,z,y_1,y_2) dx^{\mu_1}\wedge \dots \wedge dx^{\mu_5}]-W[F_5]\bigg),
\label{varform}
\end{equation}
where we can already insert the $\mathcal{O}(\gamma^0)$-solution of $F_5$.
We can interpret (\ref{varform}) as a variation of the functional 
\begin{align}
\mathcal{S}: &C^\infty(\mathcal{M}) \rightarrow \mathbb{R} \nonumber \\ &
F  \mapsto \int dx^{10} \sqrt{-g} \gamma W[F_5+ F dx^{\mu_1}\wedge \dots \wedge dx^{\mu_5}].
\end{align}
The argument, why we are allowed to assume that $F$ only depends on $u,t,z,y_1,y_2$ is the same is in the case $\mathcal{O}(\gamma^0)$, alternatively one easily verifies that
\begin{equation}
\partial_\mu \frac{\partial \mathcal{S}}{\partial \partial_\mu F}=0,\hspace{0,2 cm}\partial^2_\mu \frac{\partial \mathcal{S}}{\partial \partial_\mu^2 F}=0
\end{equation}
for $\mu \in\{x,y,y_3,y_4,y_5\}$. The results for all directions of  $\frac{\delta \mathcal{W}}{\delta F_5}$ needed to compute the EoM obtained by evaluating (\ref{EoMF5}) for the components corresponding to the right hand side of diagram \ref{diag1} and \ref{diag2} in order $\mathcal{O}(\gamma)$ can be found in the Appendix. It should be mentioned that due to the anti-self-duality of  $\frac{\delta \mathcal{W}}{\delta F_5}$ the components given in section \ref{Appendix} are all you need to compute diagrams \ref{diag1}, \ref{diag2}. The other directions can be computed from those or vanish, since we only consdier EoM, which are linearized in $A_x$.\\\\
Now let us  sketch how to solve this zoo of differential equations. One important observation is that the $x y_2 y_3 y_4$-direction of $C_4$ plays a crucial role. Considering which components of $\frac{\delta \mathcal{W}}{\delta F_5}$ are zero and which actually give contributions to (\ref{EoMF5}) shows that the argument we applied in the first section, when discussing diagram \ref{diag1}, for why the $x y_2 y_3 y_4$-direction of $C_4$ is the only non-zero one on the left hand side of diagram \ref{diag1}, doesn't change if we include $\alpha'$-corrections. Thus, diagram \ref{diag1} reduces to (\ref{diag3}) in order $\mathcal{O}(\alpha'^3)$.
\begin{equation}
\begin{xy}
  \xymatrix{
      (C_4)_{x y_2 y_4 y_5} \ar[r]^d \ar[rd]^d \ar[rdd]^d \ar[rddd]^d   &   (F_5)_{t x y_2 y_4 y_5}\ar[r]^{*} &   (*F_5)_{u y z y_ 1y_3}\ar[r]^d   & (d*F_5)_{t u y z y_ 1y_3} \ar[ddd]^=&\\
         & (F_5)_{ x z y_2 y_4 y_5}\ar[r]^{*} &   (*F_5)_{tu y y_ 1y_3} \ar[ru]^d  & \\
         &   (F_5)_{u x y_2 y_4 y_5}\ar[r]^{*} &   (*F_5)_{t  y z y_ 1y_3}\ar[ruu]^d &   \\    
     & (F_5)_{ x  y_1 y_2 y_4 y_5}\ar[r]^{*} &   (*F_5)_{tu  y z y_3} \ar[ruuu]^d & \big(d \big(* \frac{2 \gamma}{\sqrt{-g}}\frac{\delta \mathcal{W}}{\delta F_5}\big)\big)_{t u y z y_ 1y_3}\\
  }
  \label{diag3}
\end{xy}
\end{equation}
Our ansatz for  $(C_4)_{x y_2 y_4 y_5} $ will be of the form \begin{equation}
(C_4)_{x y_2 y_4 y_5} = \cos(y_1)^4\sin(2y_2)\frac{A_x + \gamma C(u,q,\omega)}{\sqrt{3}}. 
\label{ansatzC4}
\end{equation}
The $y_1,y_2$-dependence is dictated by the form of the components of $\frac{\delta \mathcal{W}}{\delta F_5}$ listed in  section \ref{Appendix} and the requirement that $\partial_{y_i}A=0$. It is possible to find a similar simplification for its dual diagram again obtained by analysing the $y_1,y_2$-dependence of  the relevant directions of $\frac{\delta \mathcal{W}}{\delta F_5}$. This has to be repeated for the remaining diagrams in order to solve the EoM for the relevant directions of $F_5$, obtained by varying the action with respect to $A_\mu$. However, this very tedious calculation can be abbreviated by an elegant shortcut, which we present in the following, see also \cite{Peeters}. We took the effort to calculate the EoMs using both methods to test our results. 
\\\\ There is also a slightly different approach to solve (\ref{EoMF5}), which relies on the observation that for every solution $F_5$ also
\begin{equation}
F_5+\gamma \tilde F 
\end{equation}
with
\begin{equation}
d\tilde F=0,\hspace{0.1cm} d\big(1-*\big)\tilde F=0,
\end{equation}
solves (\ref{EoMF5}) and fulfills that there is a four form $C_4$ with $d C_4 =F_5+\gamma \tilde F $. Let $\tilde{F_5}$ be a solution of (\ref{EoMF5}) with $d\tilde F_5=0$. Considering the de Rham-cohomology of our manifold shows that the EoM for the five form can be written as
\begin{equation}
 \bigg(-\tilde F_5+*\tilde F_5-*\frac{2\gamma}{\sqrt{-g}}\frac{\delta \mathcal{W}}{\delta F_5} \bigg)=\gamma d H_4,
\end{equation}
for some $4$-form $H_4$. Since  $\frac{\delta \mathcal{W}}{\delta F_5}  $ is anti-self dual, also $dH_4$ has to be anti-self dual. So
\begin{equation}
d\big(1-* \big)dH_4=2 ddH_4=0,
\end{equation}
such that  we can choose $\tilde{F}=-\frac{dH_4}{2}$, set 
\begin{equation}
 \tilde  F_5 = F_5 +\gamma \tilde{F}
\end{equation}
for another closed solution $F_5$ of (\ref{EoMF5}) and thus  get \begin{equation}
F_5=*\bigg(F_5-\frac{2\gamma}{\sqrt{-g}}\frac{\delta \mathcal{W}}{\delta F_5}\bigg).
\label{eqF5}
\end{equation}
The differential equation depicted in diagram (\ref{diag3}) can be deduced from the $tuyzy_3$, $uxy_2 y_4 y_5$, $txy_2 y_4 y_5$
and $zxy_2 y_4 y_5$-diection of (\ref{eqF5}). In addition it helps us to express the $tuyzy_3$-direction of $F_5$ by its $x y_1 y_2 y_4 y_5$-component and the appropriate direction of $\frac{\mathcal{W}}{\delta F_5}$. In an analogous way this links the pairs
\begin{equation}
\{(x y_1 y_2 y_3 y_5,tuyzy_4),(x y_1 y_2 y_3 y_4,tuyzy_5)\},
\end{equation}
where it turns out that up to a different $y_1,y_2$-dependence the directions $tuyzy_i$ with $i \in \{3,4,5\} $ of $F_5$ are identical, the same is valid for their dual partners. This is great news, since now we can reduce the entire coupled set of EoM for the $4$-form components and the gauge field $A_\mu$ to a rather simple system of two coupled differential equations for $A_\mu$ and the $x  y_2 y_4 y_5$-component of $C_4$. Exploiting the relations between the directions $tuyzy_i$ with $i \in \{3,4,5\} $ of the five form and the analogous ones for their dual partner gives after a tedious calculation
\begin{equation}
-\frac{1}{4\cdot 5!}\frac{\partial \sqrt{-g} F_5^2}{\partial A_x}=\frac{16 \gamma C(u,q,w)}{3 u^2}+\frac{4 (F_5)_{tuyzy_3}}{\sqrt{3}\sin(y_1)^2}.
\label{var1F5}
\end{equation}
Applying (\ref{eqF5}) gives 
\begin{align}
(F_5)_ {tuyz y_3}&=\sqrt{-g}g^{xx}g^{y_1 y_1}g^{y_2 y_2} g^{y_4 y_4}g^{y_5 y_5}\bigg(4 \sin(y_1)\cos(y_1)^3\sin(2y_2)\frac{ \gamma C(u,q,\omega)}{\sqrt{3}}\nonumber- \\& \frac{2\gamma }{\sqrt{-g}}\bigg(\frac{\mathcal{W}}{\delta F_5}\bigg)_{xy_1y_2y_4y_5} \bigg).
\end{align}
Adding up everything we can finally write down the differential equation obtained by varying with respect to $A_x$. For this purpose let us define 
\begin{equation}
A_x(u,z,t)=A_x^0(u,z,t)+\gamma A_x^1(u,z,t)
\end{equation}
\begin{equation}
A^i_x(u,z,t)=\int \frac{d^4 k}{(2 \pi)^4} \tilde{A^i}_x(u,q,w) e^{- i \omega t+i qz}
\end{equation}
with $\tilde{A}_x(u,q,w)=:(A_x)_k$, $k=(w,q)$.  
The EoM for $(A_x^1)_k$ is given by 
\begin{align}
&\partial_u^2(A^1_x)_k+\frac{2 u }{-1+u^2}\partial_u(A^1_x)_k+\frac{(\tilde q^2 (-1 + u^2) + \tilde \omega^2)}{u (-1 + u^2)^2}(A^1_x)_k+\frac{1}{(48 u^2 (-1 + u^2)^2)}(u^3 \nonumber \\ &(-9216 \tilde q^4 u^3 (-1 + u^2)  + 
   \tilde q^2 (-3900 + 73507 u^2 - 145342 u^4 + 75735 u^6) + 
    15 (520\nonumber \\ & - 1061 u^2 + 435 u^4) \tilde w^2) (A^0_x)_k -
 2 (-1 + u^2) (96 C(u, q, \omega)+ 
    u^3 (-1 + u^2) (3900 - \nonumber \\ & 23846 u^2 - 23040  \tilde q^2 u^3 + 
       675 u^4) \partial_u(A^0_x)_k))=0.
       \label{finaleq1}
\end{align}
where $\tilde{\omega}=\frac{\omega}{2 r_h}$, $\tilde{q}=\frac{q}{2 r_h}$. The coupling corrected relation between horizon radius $r_h$ and temperature $T$ is given  by $r_h=\pi T (1- \frac{265}{16}\gamma+\mathcal{O}(\gamma^2))$. If we introduce in (\ref{finaleq1}) rescaled variables  $\hat{\omega}=\frac{\bar{\omega}}{2 \pi T}$ and $\hat{q}=\frac{\bar{q}}{2 \pi T}$ we obtain a differential equation whose characteristic exponents simplify to $\pm \frac{i \hat{\omega}}{2}$ also in order $\mathcal{O}(\gamma)$. From diagram (\ref{diag3}) or the $tuyzy_3$, $uxy_2 y_4 y_5$, $txy_2 y_4 y_5$
and $zxy_2 y_4 y_5$-components of (\ref{eqF5}) we obtain the differential equation
\begin{align}
&\partial_u^2(A^1_x)_k+ \frac{2 u }{-1 + u^2}\partial_u(A^1_x)_k+\frac{\tilde q^2 (-1 + u^2) + \tilde w^2}{u (-1 + u^2)^2} (A^1_x)_k+ \frac{1}{48 u^2 (-1 + u^2)^2}\big(u^3  (-9216 \tilde q^4 u^3 \nonumber \\ & (-1 + u^2)+ 
     \tilde q^2 (-3900 + 116931 u^2 - 260414 u^4 + 147383 u^6) + 
     3 (2600 - 10969 u^2 + \nonumber \\ & 7839 u^4) \tilde w^2) (A^0_x)_k + 
  2 (24 (-2 + 2 u^2 + \tilde q^2 u (-1 + u^2) + u \tilde w^2) C(u,q,\omega) - 
     u^2 (-1 + 
        u^2) \nonumber \\ & (u (-1 + u^2) (3900 - 36702 u^2 - 32480 \tilde q^2 u^3 + 
           20895 u^4) \partial_u (A^0_x)_k - 
        24 (2 u \partial_u C(u,q,\omega)+\nonumber \\ & (-1 + u^2) 
             \partial_u^2 C(u,q,\omega))))\big)                   =0 . \label{finaleq2}
\end{align}
The boundary conditions of these EoMs are that $A_x$ and $C$, respectively the $x y_1 y_2 y_4 y_5$-component of the five form, have to be infalling at the horizon. The zeroth expansion coefficient of the near horizon expansion of $A_x/(1-u)^{-\frac{i \hat{\omega}}{2}}$ can be set to $1$, since it doesn't affect any physical observables on the boundary due to the form of (\ref{eq:Pi}). The missing condition is that $C(u,q,\omega)$ has to vanish on the boundary, which is a regular singular point of our small system of EoMs. More explicitly this can be obtained from the two different possible boundary behaviours of $C(u,q, \omega)$ given by
\begin{equation}
C(u,q, \omega) = \frac{C_{-2}}{u^2}+\mathcal{O}(u^{-1}) \hspace{0.2 cm} \text{ and } \hspace{0.2 cm} C(u,q, \omega) = u^3 C_{3}+ \mathcal{O}(u^{4}),
\end{equation}
extracted from the near boundary analysis of the differential equation obtained by subtracting (\ref{finaleq2}) from (\ref{finaleq1}).
Equation (\ref{finaleq1}) shows that the former choice would lead to a gauge field $A_x$, which diverges at the boundary. This means our missing boundary condition is that 
\begin{equation}
C_{-2}=0,
\end{equation}
which in this case implies that $C_{-1}=\dots C_{2}=0$, such that $ C(u,q, \omega) = u^3 C_{3}+ \mathcal{O}(u^{4})$.
\section{Results}
\label{sec:results}
Let us now turn to determining $\alpha'$-corrections to several observables such as the conductivity, photoemission
 rates, quasinormal mode spectra as well as in and off-equilibrium spectral densities. These were first computed in \cite{Vuorinen:2013} using the results of \cite{Hassanain:2011ce, Hassanain:2012uj, Hassanain:2011fn}, which we now argue to be incorrect. Consequently also the results for  observables, which can be found in the literature, computed with the $\gamma$-corrected EoM for gauge fields change. The differences are quite substantial and are caused by several disagreements: most importantly a missing factor $i$ in front of some components of the five form, when working in Euclidean signature, several missing terms, when computing the Hodge duals, coming from the off-diagonal elements of the metric tensor, and the fact that the five form used by the authors of the papers \cite{Hassanain:2011ce, Hassanain:2012uj, Hassanain:2011fn} did not solve it's $\alpha'$-corrected EoM. Note that in Euclidean signature there is no self duality, since the Hodge star operator squares to $-1$ there, such that self dual five forms transform to imaginary anti-self-dual forms $*F_5^{\text{E}}=-i F_5^{\text{E}}$. Continuing to work with $(1+*)F_5^{el}$ implies that the five form doesn't square to zero anymore, which means it doesn't even solve its EoM in the lowest order in $\alpha'$. Also the Lorentz-signature version of the coupling corrected five form given in \cite{Hassanain:2011ce, Hassanain:2012uj, Hassanain:2011fn} is not a solution of (\ref{EoMF5}).
 \subsection{Quasinormal modes and their coupling corrections}
 \label{sec31}
 Quasinormal modes (QNM) describe the response of the system to infinitesimal perturbations. In our case these perturbations correspond to tiny twists of the $S_5$-part of our geometry, from which we deduced the $\alpha'$-corrected differential equations (\ref{finaleq1}) and (\ref{finaleq2}) for gauge fluctuations. The spectrum of  the complex QNM-frequencies $\omega$ is the discrete spectrum of frequencies, at which the propagator of $A_x$ has poles. The negative inverse of the imaginary part of $\omega$ gives the thermalization time $\tau$, such that one can expect that increasing $\gamma$ or decreasing the 't Hooft coupling will decrease the absolute value of the imaginary part of each QNM frequency $\omega$. 
 Following   ~\cite{Son:2002} one can calculate the retarded propagator for transverse fields $\Pi_\bot$ with the help of the prescription
  \begin{equation}
  \Pi_\bot = -\frac{N^2 T^2}{8}\lim_{u \to 0}\frac{(A_x)_k'}{(A_x)_k}.
  \label{eq:Pi}
  \end{equation} 
  such that
  \begin{equation}C^{ret}_{\mu \nu}=P_{\mu \nu}^T \Pi_\bot+P^{L}_{\mu \nu}P_{\parallel},
  \end{equation}
 with $P_{\mu \nu}=\eta_{\mu \nu}-\frac{k_\mu k_\nu}{k^2}$, $P^T_{ij}=\delta_{ij}-\frac{k_i k_j}{k^2}$ and zero elsewhere, $P^L_{\mu \nu}=P_{\mu \nu}-P^T_{\mu \nu}$. Here $C^{ret}_{\mu \nu}$ denotes the retarded electromagnetic current-current correlator
 \begin{equation}
 C^{ret}_{\mu \nu} =-i \int d^4 x e^{i  kx} \theta(t)\langle [J_{\mu}^{\text{em}}(x),J_{\mu}^{\text{em}}(0) ]\rangle .
 \end{equation}
  In the following we will present several techniques with which we can extract the $\alpha'$-corrected spectra for different values of $q$ using  (\ref{finaleq1}) and (\ref{finaleq2}). 
Independent from the approach used the first information about the solutions we have to exploit is their near horizon behaviour  
\begin{align}
&A_x^0(u,\hat{q},\hat{\omega})=(1-u)^{-\frac{i \hat{\omega}}{2}}\Phi^0(u,\hat{q},\hat{\omega})\\&
A_x^1(u,\hat{q},\hat{\omega})=(1-u)^{-\frac{i \hat{\omega}}{2}}\Phi^1(u,\hat{q},\hat{\omega})\\&
C(u,\hat{q},\hat{\omega})=(1-u)^{-\frac{i \hat{\omega}}{2}}\Phi^2(u,\hat{q},\hat{\omega}).
\label{nearhor}
\end{align}
 
Let us start with an easy way to solve (\ref{finaleq1}) and (\ref{finaleq2}) with this ansatz, where the prize we pay is that the precision of our results scales more or less logarithmically with  effort. We simply expand the resulting differential equations  around the horizon and require them to hold order by order in $(1-u)$. By going to sufficiently  large orders and demanding that 
\begin{equation}
\Phi^0(0,\hat{q},\hat{\omega})+\gamma \Phi^1(0,\hat{q},\hat{\omega})=0,
\end{equation}
we can extract the $\alpha'$-corrected spectra for arbitrary values of $q$.\\ \indent Alternatively, we can apply  spectral methods  to reduce our system of differential equations to a generalized eigenvalue problem. For this purpose we use the same notation as in (\ref{nearhor}) and subtract (\ref{finaleq1}) from (\ref{finaleq2})
 to end up with a differential equation only containing $\Phi^0$ and $\Phi^2$.  We set $\Phi^2= u \tilde \Phi^2$ and \begin{equation}
 A_x(u,\hat{q},\hat{\omega})=(1-u)^{-\frac{i \hat{\omega}}{2}}\Phi(u,\hat{q},\hat{\omega}) 
 \end{equation} 
  and obtain after an expansion in $\gamma$ 
 \begin{align}
& \bigg(\partial_u^2 \tilde \Phi^2+\frac{2 + i u \hat w + i u^2 (4 i + \hat w)}{u - u^3} \partial_u \tilde \Phi^2+ \frac{1 }{4 u^2 (-1 + u) (1 + u)^2}(24 + u^2 (8 + 4 \hat q^2 - 10 i \hat w \nonumber \\ & -3 \hat w^2) + 4 u (6 + \hat q^2 - i \hat w -\hat  w^2) - 
 u^3 (-8 + 6 i \hat w + \hat w^2)) \tilde \Phi^2+\frac{u^2}{12 (-1 + u^2)}\big(-i ((3214 + \nonumber \\ &  3214 u - 5055 u^2 - 5055 u^3 + 4248 i \hat w) \hat w + 
   8 \hat q^2 (-1357 i + 295 u \hat w + u^2 (2239 i + 295 \hat w))) \Phi  \nonumber \\ & 
  -2 (-1+u^2)(-3214 - 2360 \hat q^2 u + 5055 u^2) \partial_u \Phi\big)\bigg) \gamma=\mathcal{O}(\gamma^2)
  \label{speceq1}
 \end{align}
 and 
 \begin{align}
& \partial_u^2 \Phi -\frac{i (\hat w + u (2 i + \hat w))}{-1 + u^2} \partial_u \Phi+\frac{4 \hat q^2 (1 + u) - 
 \hat w (4 \hat w + u^2 (2 i +\hat w) + u (2 i + 3\hat w))}{(4 (-1 + u) u (1 + u)^2)}\Phi+\nonumber \\ & \frac{\gamma}{48 u (-1 + u^2)}\big((-9216 \hat q^4 u^5 + 
     i (3900 u^3 - 23846 u^5 + 675 u^6 + 675 u^7 + 
        30 u^2 (130 + 313 i \hat w) \nonumber \\ & + u^4 (-23846 - 6525 i \hat w) + 
        1590 i \hat w) \hat w + 
     \hat q^2 (1590 + 3900 u^2 - 69607 u^4 + 45 u^6 (1683 -\nonumber \\ &  512 i \hat w) - 
        23040 i u^5 \hat w)) \Phi - 
  2 (96 \tilde{\Phi}^2 +  u^2(-1 + u^2) (3900 - 23846 u^2 - 23040 \hat q^2 u^3 + \nonumber \\ & 
        675 u^4)\partial_u \Phi)\big)=\mathcal{O}(\gamma^2).
          \label{speceq2}
 \end{align}
 We can solve $\Phi^0$ for a given value of $\hat{\omega}$
at a certain $\hat q$ using spectral methods almost up to arbitrary numerical precision, due to the simplicity of the order $\mathcal{O}(\gamma^0)$-EoM. It would even be possible to find analytic solutions in the lowest order in $\gamma$, but for our purposes an approximation by Chebyshev-cardinal functions is sufficient, if we choose the order sufficiently high or the Gauss-Lobatto grid sufficiently dense. 
 We also approximate $\Phi$ and $\tilde \Phi^2$ in the following by  a truncated expansion in cardinal functions on a Gauss-Lobatto grid
\begin{equation}
\big\{-\cos(\frac{\pi n}{M})\big\}_{n \in \{0,\dots,M\}}
\label{grid1}
\end{equation} on the interval $[-1,1]$ for $2 u-1$, $u \in [0,1]$, respectively with a grid \begin{equation}
\bigg\{\frac{1-\cos(\frac{\pi n}{M}) }{2}\bigg\}_{n \in \{0,\dots,M\}}
\end{equation} on the interval $[0,1]$. More explicitly we set for a certain value of $\hat q$
\begin{equation}
\Psi(u,\hat \omega )= \sum_{i=0}^M a_i^{\Psi}(\hat{\omega}) c(i,2u-1),
\end{equation}
with $c(i,x)$, $x  \in [-1,1]$ being the $i$-th cardinal function for the grid (\ref{grid1}) and $\Psi \in \{\Phi\text{ , }\tilde{\Phi}^2\}$. Now we can bring (\ref{speceq1}) and (\ref{speceq2}) into the form of a generalized eigenvalue problem for $\hat{\omega}$, if we  truncate the differential equations after the first order in $\gamma$. In the next step we also put $\gamma$ on a appropriate Gauss-Lobatto grid and solve the generalized eigenvalue problem for each grid point. At $\gamma=0$ the slopes of the resulting curves of partially resummed poles for different values of $\gamma$ in the complex plane  gives us the $\mathcal{O}(\gamma^1)$-coefficient to the corresponding $\lambda=\infty$-modes. For the first modes these curves are depicted in figure \ref{curves}. By going to sufficiently dense grids we obtain identical values as with the simpler  Frobenius-method discussed above. 
\begin{figure}
\includegraphics[scale=0.95]{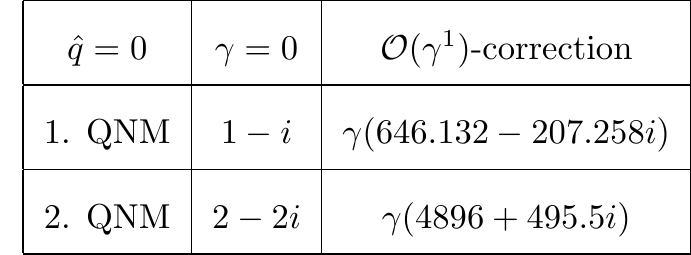}
\includegraphics[scale=0.95]{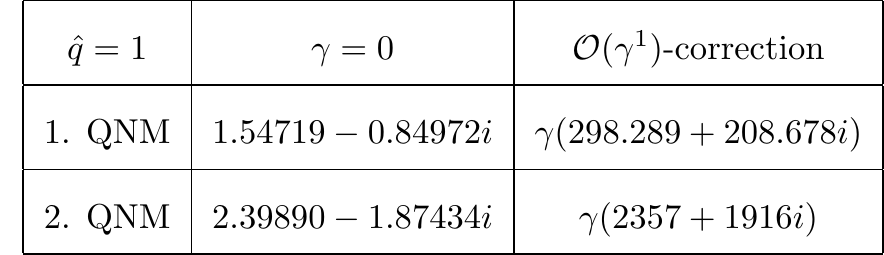}
\caption{The first two QNM frequencies at $q=2 \pi T$ (right) and $q=0$ (left) normalized by $2 \pi T$ and their $\mathcal{O}(\gamma)$-corrections, which turn out to be more than one order of magnitude smaller then found in \cite{Vuorinen:2013}, which was based on the EoM derived in \cite{Hassanain:2011ce, Hassanain:2012uj, Hassanain:2011fn}.}
\end{figure}
\begin{figure}
\includegraphics[scale=0.9]{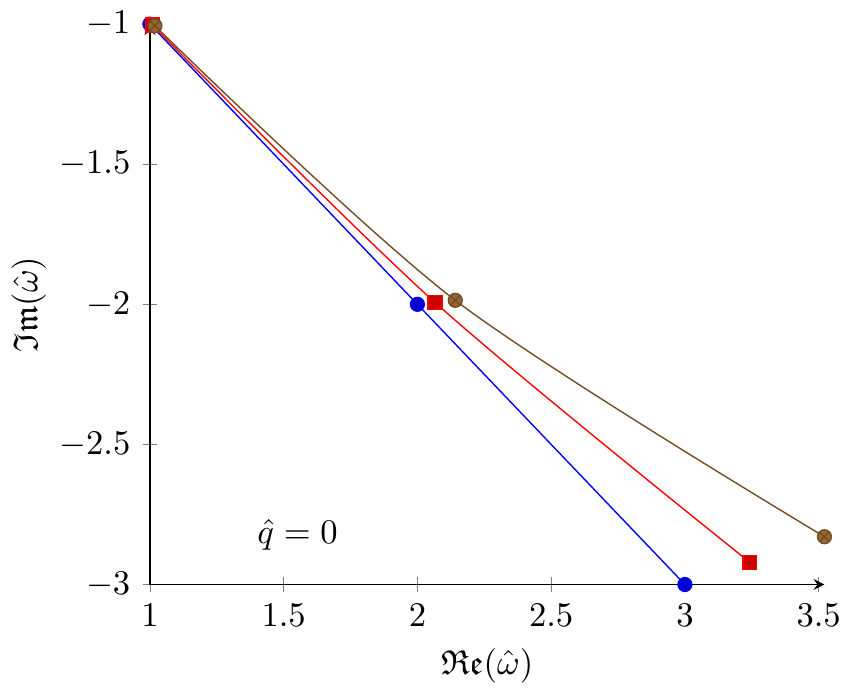}
\includegraphics[scale=0.9]{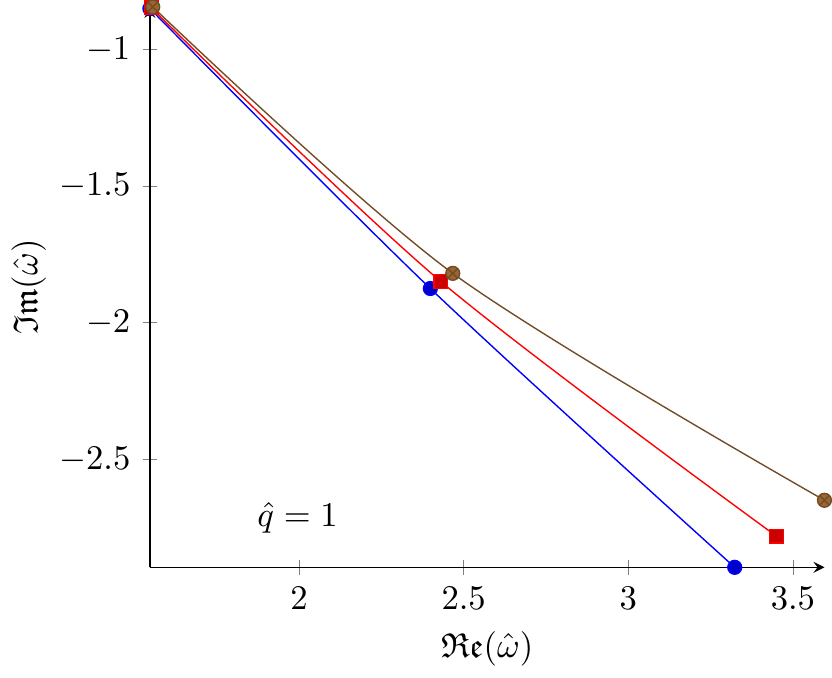}
\caption{The first QNM frequencies at $q=2 \pi T$ (right) and $q=0$ (left) normalized by $2 \pi T$ for $\lambda= \infty$ (blue) and their $\mathcal{O}(\gamma)$-corrections for $\lambda=500$ (red) and $\lambda =300$ (brown).}
\end{figure}
\FloatBarrier
\subsection{Finite coupling corrections to the plasma conductivity and photoemission rate}
  In order to compute the spectral density respectively the photoemission rate and its finite coupling corrections from our transverse field $A_x$  we simply need the retarded  Greens function, or more precisely its imaginary part. The transverse components of the spectral density are given by \cite{Vuorinen:2013, CaronHuot:2006te}
  \begin{equation}
  \chi_{\bot}= -4 \text{Im}(\Pi_{\bot}).
  \label{eq:chi2}
  \end{equation}
  From the low energy regime respectively the first order coefficient of (\ref{eq:chi2}) in $\hat{q}$ with lightlike momentum  we can immediately read off the correction to the conductivity. The correction factor to the  differential photon production  rate can be computed via the relation between the spectral function $\chi$ and the Wightman function \cite{Vuorinen:2013}
  \begin{equation}
\Pi_{\mu \nu}^<=n(k)\chi_{\mu \nu}(k),
\end{equation}   
with $n(k)=1/(e^{\frac{k}{T}}-1)$, such that
\begin{equation}
\frac{d \Gamma}{d k}=\frac{\alpha_{\text{em}}n(k)}{\pi}k \chi_\mu^\mu.
\end{equation}
To obtain the low energy limit of (\ref{eq:chi2}), more specifically the finite coupling correction to the conductivity, we only have to solve (\ref{finaleq1}) and (\ref{finaleq2}) to order $\mathcal{O}(\gamma)$ and $\mathcal{O}(\hat \omega)$. In this case the solution for $C(u,\hat q,\hat{\omega})$ is simply
   \begin{equation}
   C(u,\hat q,\hat{\omega})=\bigg(\frac{95 u^3}{8} - \frac{959 u^5}{24} + \frac{337 u^7}{12}\bigg)\partial_u A^0_x+\mathcal{O}(\hat \omega^2),
\end{equation}      
which means that our EoM for $A_x$ simplifies drastically to 
\begin{equation}
\partial_u^2 A_x+\frac{u}{8 (-1 + u^2)}\big( 16 \partial_u A_x + \gamma (920 - 7970 u^2 + 7275 u^4 - 225 u^6)\partial_u A_x\big)= \mathcal{O}(\gamma^2,\hat{\omega}^2)
\label{eqorderw}
\end{equation}
 Here it suffices to apply Frobenius methods, since after only a couple of orders in $(1-u)$, we obtain  stable results. We expand the functions $A_x$ at the horizon
\begin{equation}
  A_x=(1-u)^{-\frac{i \hat{\omega}}{2}}\sum_{i=1}^K( a_i (1-u)^i+\gamma b_i (1-u)^i),
  \label{eq:Psi}
  \end{equation}
  with $K$ sufficiently large.
  Inserting this ansatz into (\ref{eqorderw}) and solving the resulting equation order by order in $(1-u)$ as well as order by order in $\gamma$ and only up to order $\mathcal{O}(\hat \omega)$  gives us a low energy approximation of the solution of $A_x$ near the horizon. We continue this computation  until we have reached a $K$ for which the numerical results for  the conductivity and its $\gamma$ correction stabilize. We counterchecked our findings  by calculating $A_x$ from (\ref{finaleq1}) and (\ref{finaleq2}) with the help of spectral methods and took the low energy limit of (\ref{eq:chi2}). For the spectral density in the low energy regime and lightlike momenta we find 
  \begin{equation}
  \chi_\bot^{\omega=q}= \frac{N^2 T^2}{2}\bigg((1+125 \gamma  ) \hat{q} +\mathcal{O}(\hat{q}^2) \bigg)+\mathcal{O}(\gamma^2).
  \end{equation} This means that the conductivity $\sigma$  gets a $\gamma$-correction factor of $(1+ 125 \gamma)$. This is identical to the finite coupling correction factor for the  photoemission rate at  $1 \gg \omega $,  which coincides with the expectations of \cite{CaronHuot:2006te} for the low frequency limit, which predicted a growing behaviour for decreasing  't Hooft coupling in this regime. \\ \indent Let us now turn to the large $\omega$ calculation. This is interesting, since originally the authors of  \cite{CaronHuot:2006te} expected the photoemission rate to decrease with decreasing $\lambda$. However, the authors in \cite{Hassanain:2011ce, Hassanain:2012uj, Hassanain:2011fn} found a correction factor of $(1+5\gamma )$, which would indicate the contrary behaviour. Thus we want to see if this behaviour still holds, when using the correct EoM.  We choose to determine the functions $\Phi^0$, $\Phi^1$, $\Phi^2$ as an approximation in cardinal functions and compute the large $\omega$ limit numerically in the zero-virtuality case $\omega=q$. By using sufficiently large Gauss-Lobatto grids we find the following large-$q$ behaviour
  \begin{equation}
    \chi_\bot^{\omega=q, q \gg 1}= \frac{N^2 T^2}{4}\frac{ 3^{5/6}\Gamma(\frac{2}{3})}{\Gamma(\frac{1}{3})}\bigg((1-80.39 \gamma  ) \hat{q}^{2/3} +\dots \bigg)+\mathcal{O}(\gamma^2),
  \end{equation}  
  where dots stand for terms of order $\hat{q}^\alpha$ with $ \alpha < \frac{2}{3}$. In the same way as before we can read off the correction factor to the photoemission rate  from this result.\\
  \indent We now want to compare our small and large $\omega$ limits with the analogous ones for the spectral density in the spin-$2$ channel. A quite similar calculation there (as obtained in   ~\cite{Buchel:2008sh}) gives  for $1 \gg \omega$ a correction factor  $(1+135 \gamma)$. We performed  a numerical large $\omega$ analysis of the spectral density in the lightlike case also in this channel and obtained a  correction factor of $(1- \frac{290}{3} \gamma )$ there (actually our result was of the form $(1-\gamma 96.66666\dots7)$ with sufficiently many digits that we can write $\frac{290}{3}$). To sum up we find a quite similar behaviour of the $\gamma$-corrected spectral density and photoemission rates in the spin $1$ and spin $2$ channel, whose sign of the correction factors coincide in both limits with the intuitive expectations, respectively the expectations of  ~\cite{CaronHuot:2006te}.
  \subsection{Finite coupling corrections to the off-equilibrium spectral density}
  Let us finally turn to determining the $\gamma$-corrected on-shell photoemission spectrum in the off-equilibrium case. For this purpose we consider the simplified setting of   ~\cite{Vuorinen:2013}. There, the authors consider an infinitely thin shell, collapsing towards its horizon in the static coupling-corrected $AdS_5$-background. It is assumed that the shell is collapsing so slowly that its radial motion can be neglected. Let us start with the $\gamma=0$ case. The motivation for the form of the metric we use is given by Birkhoff's theorem, stating that outside of the shell the solution for the Einstein equations is the $AdS$-Schwarzschild metric, whereas inside of the shell we have a pure $AdS$-space. This implies
\begin{equation}
ds^2=\frac{r_h^2}{u^2}\bigg(f(u)dt^2+dx^2+dy^2+dz^2 \bigg)+\frac{1}{4u^2f(u)}du^2
\end{equation}
with 
\begin{equation}
f(u)=\begin{cases} f_-(u)=1& \text{if } u> u_s\\f_+(u)=1-u^2& \text{if } u_s>u.\end{cases}
\end{equation}
and $u_s=\frac{r_h^2}{r_s^2}$, where $r_s$ is the radial position of the shell. Requiring that the metric, solutions for fluctuations etc. are continuous at the position of the shell will give us junction or matching conditions.  In order $\mathcal{O}(\gamma )$  the metric outside of the shell will be  (\ref{eq:metricc}), whereas inside of the shell we have no coupling corrections at all ~\cite{Buchel:2008sh}.  From this we can immediately read off the matching condition for the  frequency
\begin{equation}
\hat{\omega}_+= \hat{\omega}_- \sqrt{U(u_s)u_s},
\end{equation}  
by comparing the prefactors of $dt^2$ in the line elements inside of and outside of the shell.
Here and in the following subindices $_+$ denote quantities outside of the shell and subindices $_-$ inside of the shell. From the requirement that the metric has to be continuous, it follows that $dx_+=dx_-$ and the same for $y$ and $z$. The calculation we perform in the following is identical to the one, where we require the continuity of  the gauge invariant combination $\omega A_x$. For everyone, who doesn't want to  work with $A_x$ instead of $E=\omega A_x $, can think of $A$, which is the notation for the transverse gauge fields we use in the following, as $A=\omega A_x$.
Since we have $t_-=\sqrt{U(u_s)u_s}t_+$ and since we require $A$ to be continuous at the shell position, we obtain
\begin{equation}
A_+(u,z,t)\big|_{u=u_s}=A_-(u,z,t\sqrt{U(u_s)u_s})\big|_{u=u_s},
\end{equation}
thus 
\begin{align}
&\bar A_+(u,q,\omega_+)\big|_{u=u_s}=\int dt e^{- i \omega_+ t}\tilde  A_+(u,q,t)\big|_{u=u_s}=\int \frac{dt}{\sqrt{U(u_s)u_s}} e^{- i \omega_- t}\tilde  A_-(u,q,t)\big|_{u=u_s}\nonumber \\& =\frac{\bar A_-(u,q,\omega_-)}{\sqrt{U(u_s)u_s}}\bigg|_{u=u_s},
\label{cont1}
\end{align} 
where functions with tilde $\tilde A_{\pm}$ and  $\bar A_{\pm}$  stand for the Fourier transformed ones. In the following we will write simply $A$ for $\bar A$ and $\tilde A$  and indicate to which functional space $A$ belongs by the variables it depends on. \\ \indent  
For derivatives in $t$-direction things are similarly easy. We have
\begin{align}
\partial_{t_-}A_-(u_s,z,t_-)=\frac{1}{\sqrt{U(u_s)u_s}} \partial_{t_+}A_-(u_s,z,t_+ \sqrt{U(u_s)u_s})=\frac{1}{\sqrt{U(u_s)u_s}}\partial_{t_+}A_+(u_s,z,t_+).
\label{cont2}
\end{align}
For derivatives in $u$-direction the junction condition turns out to be slightly more difficult to derive. Inside of the shell the EoM for $A_-$ is given by
\begin{equation}
\partial_u^2 A_-(u,\hat q,\hat \omega)+\bigg(1+\frac{265}{8}\gamma \bigg)\frac{\hat \omega^2- \hat{q}^2}{ u}A_-(u,\hat q,\hat \omega)=0,
\end{equation}
where the $\gamma$-correction merely arises due to the modified relation between $r_h$ and $T$.
Outside of the shell the  $\mathcal{O}(\gamma^0)$ part of $A$ is a solution to (\ref{eq0}), the $\mathcal{O}(\gamma^1)$ part  is a solution to (\ref{finaleq1}).  From the continuity requirement of $C(u,q,\omega)$, or the $x y_2 y_4 y_5$-component of $C_4$, we can derive a relation for $(C_4)_{x y_2 y_4 y_5}$ analogous to (\ref{cont1}). Inside of the shell we have 
\begin{equation}
((C_4)_{x y_2 y_4 y_5})_-= \cos(y_1)^4 \sin(2 y_2) \frac{A_-}{\sqrt{3}},
\end{equation}
such that at $u=u_s$ 
\begin{equation}
((C_4)_{x y_2 y_4 y_5})_+=\cos(y_1)^4 \sin(2 y_2) \frac{A_+}{\sqrt{3}}\bigg|_{u=u_s}, 
\end{equation}
which means that the contributions of $C_4$ to (\ref{finaleq1}) vanish on the surface of the  shell. Therefore, at $u =u_s$ we can write the EoM for $A_+$ as 
\begin{align}&0= \nonumber \\
&\partial_u^2 A_+(u,\hat q,\hat{\omega}_+)\big|_{u_s}+f_+^1(u_s,\hat q,\hat{\omega}_+,\gamma)\partial_u A_+(u,\hat q,\hat{\omega}_+)\big|_{u_s}+f_+^2(u_s,\hat q,\hat{\omega}_+,\gamma)A_+(u,q,\hat{\omega}_+)\big|_{u_s},
\end{align} 
whereas for $A_-$ we have
\begin{equation}
\partial_u^2 A_-(u,\hat q,\hat \omega_-)\big|_{u_s}+f_-(u_s,\hat q,\hat{\omega}_-,\gamma)A_-(u,\hat q,\hat \omega_-)\big|_{u_s}=0,
\end{equation}
with $f_+^1$, $ f_+^2$ and $f_-$ chosen appropriately.
Using (\ref{cont2}) gives
\begin{align}
f_-(u_s,\hat q,\hat{\omega}_-,\gamma)A_-(u,\hat q,\hat \omega_-)\big|_{u_s}\rightarrow \bigg(1+\frac{265}{8}\gamma \bigg)\frac{\hat \omega_+^2- \hat{q}^2f_m^2}{ u f_m^2}A_+(u,\hat q,\hat \omega_+)\big|_{u_s},
\end{align}
with $f_m=\sqrt{U(u_s)u_s}$. We now  perform a coordinate transformation such that the EoM inside  and outside of the shell are of the same shape. For this purpose we choose $\tilde{u}(u)$ such that 
\begin{equation}
\frac{d^2\tilde{u}}{du^2} +\frac{d \tilde{u}}{du} f_+^1(\hat{\omega}_-,u,\gamma,q)=0,
\end{equation} outside of  the shell and $\tilde{u}=u$ inside of it. The EoM  in this new coordinate reads outside of the shell
\begin{equation}
0=\partial_{\tilde{u}}^2 A_+(u(\tilde{u}),q,\hat{\omega}_+,\gamma)+\tilde{f}_+(u(\tilde{u}),\hat q,\hat{\omega}_+,\gamma)A_+(u(\tilde{u}),q,\hat{\omega}_+,\gamma),
\end{equation}
with \begin{equation}
\tilde{f}_+(u(\tilde{u}),\hat q,\hat{\omega}_+,\gamma):=\Big( \frac{du}{d\tilde{u}}\Big)^{2}f_+^2(u(\tilde{u}),\hat q,\hat{\omega}_+,\gamma).
\end{equation}
We can read off the junction condition for $\partial_uA_{\pm}$, by considering
\begin{equation}
\partial_{\tilde{u}}A_--\partial_{\tilde{u}}A_+=\lim_{\epsilon \to 0}\int_{u_s-\epsilon}^{u_s+\epsilon}\partial_{\tilde{u}}^2A=0,
\end{equation} which can be achieved by choosing
\begin{equation}
\bigg( \frac{d\tilde{u}}{du}\bigg)\bigg|_{u=u_s}=\sqrt{\big(u_s f_m^2\big)\frac{f_+^2(u_s,\hat q,\hat{\omega}_+,\gamma)}{\hat \omega_+^2- \hat{q}^2f_m^2}}\bigg(1-\frac{265}{16}\gamma  \bigg).
\end{equation}
By an analogous computation as in (\ref{cont1}) we obtain 
\begin{equation}
\partial_u A_-(u,\hat{q},\hat{\omega}_-)\bigg|_{u_s}=f_m\sqrt{\big(u_s f_m^2\big)\frac{f_+^2(u_s,\hat q,\hat{\omega}_+,\gamma)}{\hat \omega_+^2- \hat{q}^2f_m^2}}\bigg(1-\frac{265}{16}\gamma  \bigg)\partial_u A_+(u,\hat{q},\hat{\omega}_+)\bigg|_{u_s}.
\end{equation}In the lightlike case this explicitly means 
\begin{align}&
\sqrt{U(u_s)u_s }\bigg( \sqrt{1-u_s^2} - 
 \frac{1}{96} \big(u_s^2 \sqrt{1 - u_s^2} (53692 - 136807 u_s^2 + 75735 u_s^4 + 9216 u_s \hat w^2\nonumber \\ & - 
      9216 u_s^3 \hat w^2)\big) \gamma\bigg)\partial_u A_+(u,\hat{\omega}_+)\bigg|_{u_s}=f_m^\gamma f_m \partial_u A_+(u,\hat{\omega}_+)\bigg|_{u_s}=\partial_u A_-(u,\hat{\omega}_-)\bigg|_{u_s},
\end{align}
with 
\begin{align}
f_m^\gamma= \sqrt{1 - u_s^2} \bigg(1 - 
 \frac{1}{96} \big(u_s^2  (53692 - 136807 u_s^2 + 75735 u_s^4 + 9216 u_s \hat w^2 - 
      9216 u_s^3 \hat w^2)\big)\gamma \bigg).
\end{align}
Outside of the shell we have both ingoing and outgoing wave solutions, so that we write
\begin{equation}
A_+(u,q=\omega)=c_{\text{in}}A_{\text{in}}(u,q=\omega)+c_{\text{out}}A_{\text{out}}(u,q=\omega),
\end{equation}whereas inside of the shell we only have ingoing modes.
From the matching conditions deduced above one obtains the following relation
\begin{equation}
\frac{c_{\text{out}}}{c_{\text{in}}}=-\frac{f_m^\gamma  A_-\partial_uA_{\text{in}}-A_{\text{in}}\partial_uA_-}{f_m^\gamma  A_- \partial_u A_{\text{out}}- A_{\text{out}}\partial_uA_-}\bigg|_{u_s}.
\label{cpm}
\end{equation} At this point we can perform a non trivial check of our calculation, since obviously $\frac{c_{\text{out}}}{c_{\text{in}}} \rightarrow 0$ for $u_s \rightarrow 1$ should hold.  The outgoing  solution of $A$  inside of the shell for general virtuality, expressed by $w_+=:\omega$ is
\begin{align}
&A_-(u,\hat q,\hat{\omega})=\nonumber \\ &\sqrt{u}\bigg(J_1\Big(2\hat{\omega}\Big(1+\frac{265}{16}\gamma \Big) \sqrt{c(u_s,q/\omega) u} \Big)+iY_1\Big(2\hat{\omega}\Big(1+\frac{265}{16}\gamma \Big) \sqrt{c(u_s,q/\omega) u} \Big) \bigg),
\end{align}
with
\begin{equation}
c(u_s,q/\omega)=\bigg(\frac{1}{U(u_s)u_s}-\frac{q^2}{ \omega^2}  \bigg).
\end{equation}
Setting $q=\omega$, inserting the solution above into (\ref{cpm}) and taking the limit $u_s \to 1$ actually gives $\frac{c_{\text{out}}}{c_{\text{in}}} \to 0$ both in order $\mathcal{O}(\gamma^0)$ and $\mathcal{O}(\gamma^1)$ as expected. \\\indent
The coupling corrected off-equilibrium spectral density is given by \cite{Vuorinen:2013}
\begin{equation}
\chi(\hat \omega,u_s)= \frac{N^2 T^2}{2}(1-\frac{265}{8}\gamma)\text{Im} \bigg(\frac{\partial_u A_+}{A_+} \bigg)\bigg|_{u=0},
\end{equation}
with \begin{equation}
\text{Im} \bigg(\frac{A_+'}{A_+} \bigg)\bigg|_{u=0}=\text{Im} \bigg(\frac{\frac{c_{\text{out}}}{c_{\text{in}}} \partial_u A_\text{out}+\partial_u A_\text{in}}{\frac{c_{\text{out}}}{c_{\text{in}}} A_\text{out}+A_\text{in}} \bigg)\bigg|_{u=0}.
\end{equation}
As in \cite{Vuorinen:2013} we compare the cases $u_s=1$ and $u_s=\frac{r_h^2}{r_s^2}$ with $r_s>r_h$ by calculating the quantity
\begin{equation}
R(\hat{\omega},u_s)= \frac{\chi(\hat{\omega},u_s)-\chi(\hat{\omega},1)}{\chi(\hat{\omega},1)}.
\end{equation}
\begin{figure}
\centering
\hspace*{-5em}
\includegraphics[scale=0.9]{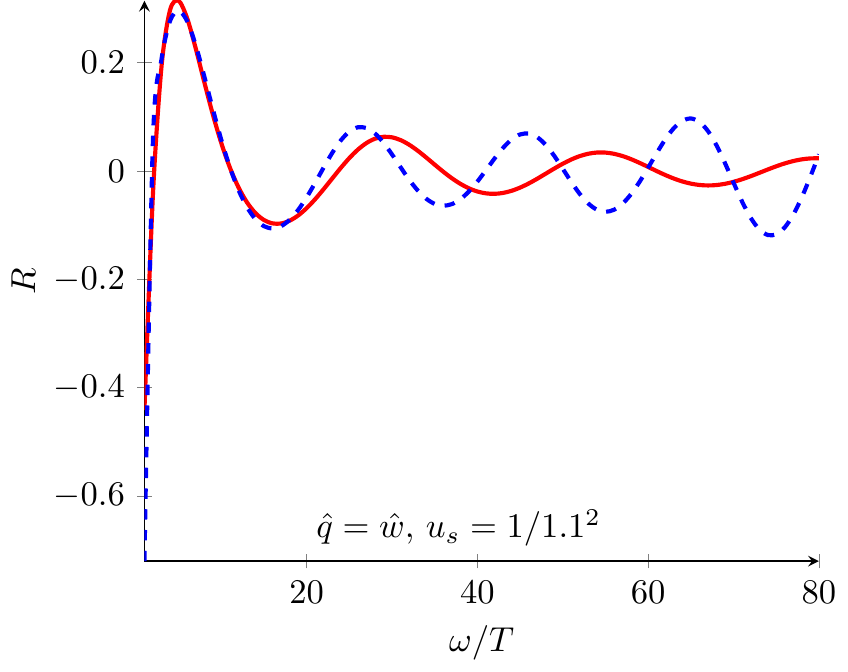}
\includegraphics[scale=0.9]{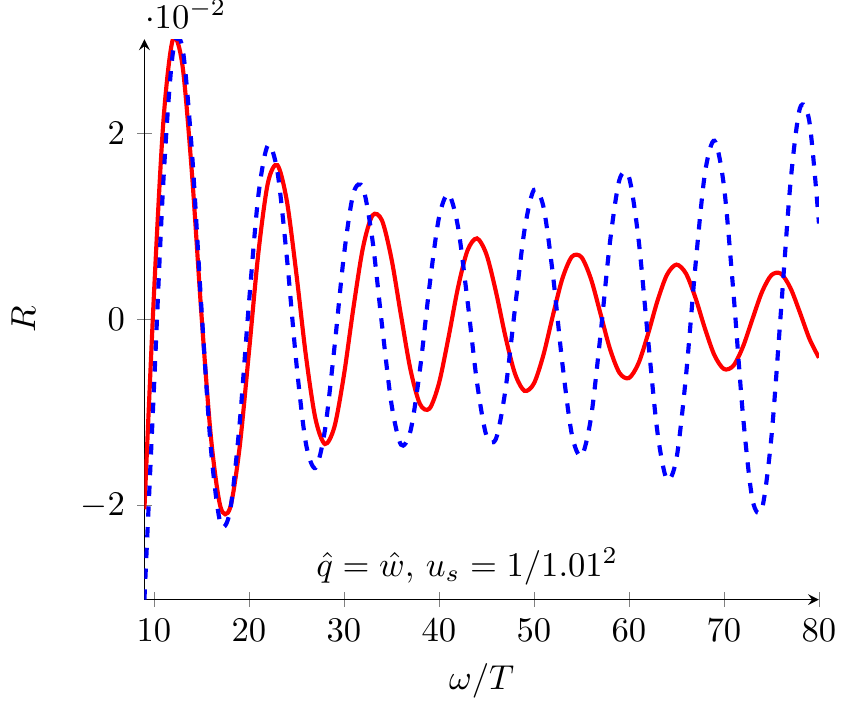}
\hspace*{-4em}
\caption
    {%
    The function $R_\bot$ plotted for $r_h=1.1$ on the left side and $r_h=1.01$ on the right side. In both pictures the solid red line represents the $\lambda=\infty$ limit, whereas the dashed blue line shows the $\mathcal{O}(\gamma^1)$ corrected results at $\lambda=300$.
    }
\label{figspec}
\end{figure}
\FloatBarrier
Figures \ref{figspec} demonstrate that even with the new EoM for transverse gauge fluctuations including $\gamma$-corrections, the results  of \cite{Vuorinen:2013} regarding the behaviour of the off-equilibrium spectral densities didn't change on a qualitative level. 
\subsection{A partial resummation of the expansion in $\gamma$}
\begin{figure}
\includegraphics[scale=0.9]{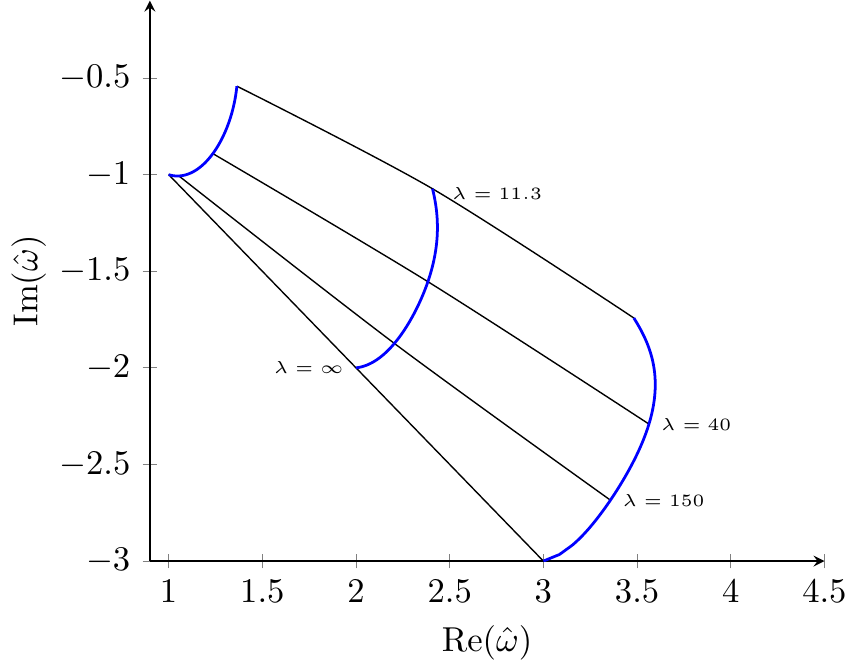}
\includegraphics[scale=0.9]{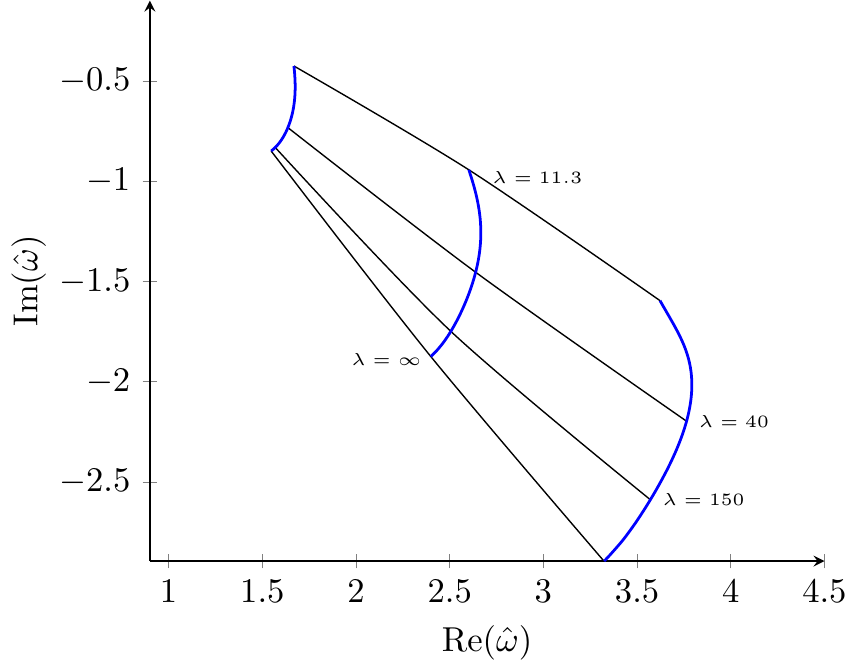}
  \caption{The flow of the first $3$ QNM frequencies, normalized by $2 \pi T$, with the 't Hooft coupling between $\lambda= \infty$ and $\lambda=11.3 \approx 4 \pi \alpha_s N$, with $N=3$ and $\alpha_s=0.3$ computed in the resummation scheme \cite{ourpaper} with $\hat{q}=0$ (left) and $\hat{q}=1$ (right). The slopes of the curves at $\gamma=0$ give the first order corrections \ref{sec31}.
    }
    \label{curves}
\end{figure}

\FloatBarrier
So far we have considered corrections to several observables, related to $\gamma$-corrected gauge fields on the gravity side or the current-current correlator on the field theory side. We are clearly not allowed to go to very  small values of $\lambda$, if we only consider $\alpha'^3$-corrections and ignore higher ones, since e.g. the QNM-spectrum will unavoidably bend upwards and eventually, at a $\lambda$-value that is sufficiently small, cross the real axis. However, there is a technique, which would in principle allow us to go to arbitrary small values for $\lambda$, without witnessing unphysical behaviour like poles with positive imaginary part. The idea is to treat the $\mathcal{O}( \gamma)$ differential equation for $A_x$ as its complete EoM and calculate exactly in $\gamma$ henceforth \cite{ourpaper}. This  is equivalent to computing all higher order corrections to certain quantities like QNM, which arise only from the $\mathcal{O}(\gamma)$-part of its EoM and resum those contributions. The results obtained hereby should be interpreted carefully. In no way is it guaranteed that we get even close to the real values at very small $\lambda$, but since even higher derivative terms to the type IIb action are not explicitly known so far, this procedure delivers the best results for small $\lambda$, which are  available at this point. \\ \indent For the QNM-spectra the calculation was already explained in section \ref{sec31}. For a given value of $\hat{q}$ and $\gamma$ using spectral methods we reduce (\ref{finaleq1}) and (\ref{finaleq2}) to generalized Eigenvalue problem for $\omega$  and repeat the calculation for points of a sufficiently dense grid, on which we have put $\gamma$. The endpoints of this curve are  $\gamma=0$ and  $\gamma =\frac{\zeta(3)}{8}(11.3)^{-\frac{3}{2}}$, the latter of  which corresponds to the value of $\lambda$ naively obtained from the QCD-limit $\alpha_s=0.3$ and $N=3$. For $\hat{q}=1$ and $\hat{q}=0$ these results are displayed in figure \ref{curves}. Technically  it is possible to go to arbitrary small values of $\lambda$, regarding the curves in figure \ref{curves}. However, the exact size of the $ \lambda$-interval in which the resummed poles still are reliable results is unclear, such that going to $\lambda=11.3$ already is quite daring. Throwing all caution aboard and analyzing the resummed spectrum for values of $1 \gg \lambda $ makes the poles align near the real axis with very small but still negative imaginary part.
\begin{figure}
\includegraphics[scale=0.9]{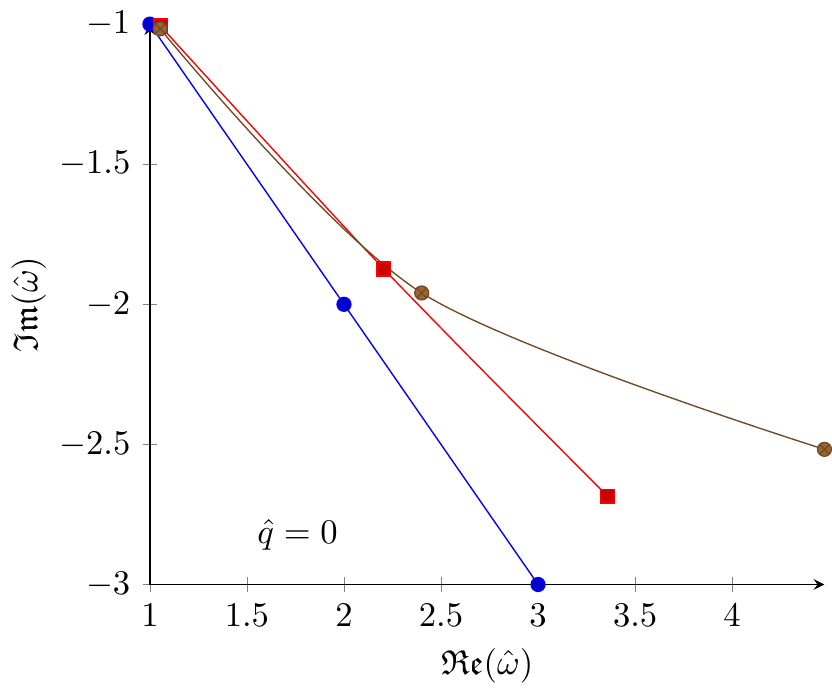}
\includegraphics[scale=0.9]{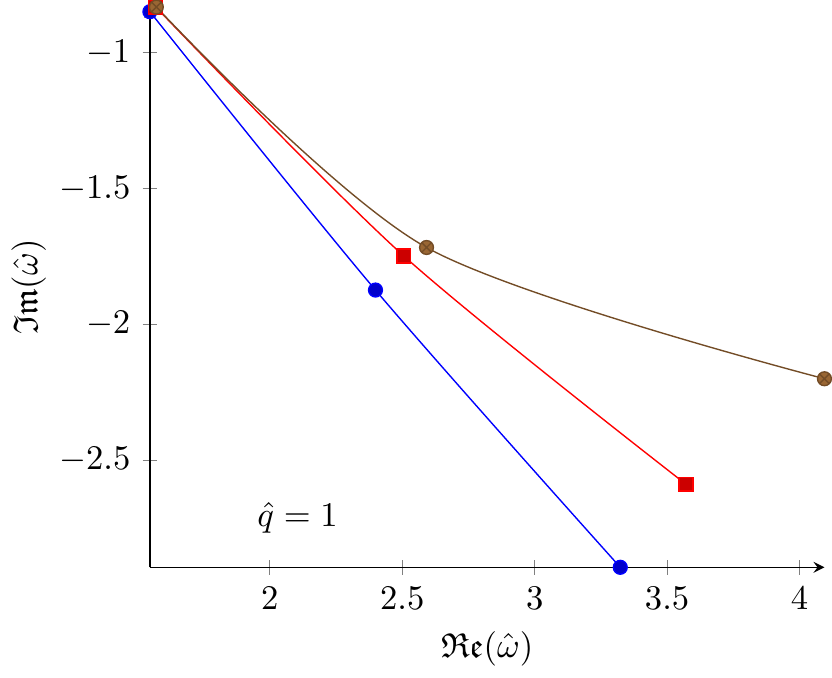}
  \caption{The first QNM frequencies at $q=2 \pi T$ (right) and $q=0$ (left) normalized by $2 \pi T$ for $\lambda= \infty$ (blue) and their $\mathcal{O}(\gamma)$-corrections for $\lambda=150$ (brown) and the resummed poles also taken at $\lambda =150$ (red).
    }
    \label{polesresum}
\end{figure}
\FloatBarrier
\section{A surprising observation}
\label{sec:why}
In section \ref{sec3} we derived the higher derivative correction to the EoM of gauge fields $A_x$. Everything followed strictly from the $\gamma$-corrected type IIb action. Now we will try a different approach, which is calculationally much easier but not mathematically well justified without further insight. Surprisingly, however, it gives identical results regarding the $\mathcal{O}(\gamma)$-corrections to the conductivity, the QNM, the photoemission rate, etc. It should be noted that the $\gamma$-correction to the off-equilibrium spectral density, see figure \ref{figspec}, differ from the actual results obtained in the previous section. This suggests that this prescription might only be valid for in-equilibrium quantities, which can be computed from the gravitational propagator (\ref{eq:Pi}). Still even such a limited validity is not understood.
 There are good reasons to assume that the prescription of inserting twice the contribution of the magnetic part of $F_5$ to $\frac{F_5^2}{4 \cdot 5!}$ back into the action and varying with respect to $A_x$ hereafter, which is valid in order $\mathcal{O}(\gamma^0)$, is simply wrong in order $\mathcal{O}(\gamma^1)$. First of all the five form $F_5$ loses its self duality in order $\mathcal{O}(\gamma^1)$. The "doubling" of the contribution of the magnetic part in the action comes from exactly there. Second and even worse, we now have further highly non-trivial terms $\gamma W$ containing $(1+*)F_5$ and derivatives  thereof. Therefore it is not only highly doubtful whether the prescription regarding the five form, which we are used to in order $\mathcal{O}(\gamma^0)$, is still working. It is not even clear how exactly it should look like.
Still one intuitive ansatz one could try is the following: 
Take the solution of the magnetic part of the five form obtained in order $\mathcal{O}(\gamma^0)$ and look at its dependence on the metric components $g^{\mu \nu}$ and $A_\mu$. Insert the $\alpha'$-corrected background metric given in (\ref{metric}). Choose the $L(u)$-prefactor of certain components of your resulting form  in such a way that 
\begin{equation}
d F^{mag}= \mathcal{O}(\gamma^2).
\end{equation}
Explicitely this means 
\begin{align}&
(F_5^{mag})^0= 4 \sqrt{\det(g_{S_5})}dy_1\wedge dy_2 \wedge  dy_3 \wedge dy_4 \wedge dy_5+\frac{4}{L(u)^5}\sqrt{|\det(g_{10})|}\sqrt{|\det(g_5)|}\nonumber \\ & \bigg(g_{10}^{t t}g_{10}^{u u}g_{10}^{yy}g_{10}^{x y_3}g_{10}^{z  z} dy_1 \wedge dy_2 \wedge dx \wedge dy_4 \wedge dy_5 +g_{10}^{t t}g_{10}^{u u}g_{10}^{yy}g_{10}^{x y_4}g_{10}^{zz } dy_1 \wedge dy_2   \nonumber \\ &\wedge dy_3\wedge dx \wedge dy_5+g_{10}^{t  t}g_{10}^{u  u}g_{10}^{ y y }g_{10}^{x y_{5}}g_{10}^{ z z} dy_1 \wedge dy_2 \wedge dy_3 \wedge dy_4 \wedge dx\bigg).
\end{align}
\begin{align}
\frac{(F_5^{mag})^1_{ux}}{L(u)^4}&= -\sqrt{\det(g_{S_5})}\big(\sin(y_1)\cos(y_1)g_{10}^{y_1 y_1}g_{10}^{y_3 y_3} du \wedge dx \wedge dy_2 \wedge dy_5 \wedge dy_4 +\nonumber \\ &  \cos(y_1)^2\sin(y_2)\cos(y_2)g_{10}^{y_2 y_2}g_{10}^{y_4 y_4} du \wedge dx \wedge dy_1 \wedge dy_5 \wedge dy_3-\sin(y_1)\times \nonumber \\ &\cos(y_1)\sin(y_2)^2g_{10}^{y_1 y_1}g_{10}^{y_4 y_4} du \wedge dx \wedge dy_2 \wedge dy_3 \wedge dy_5-\cos(y_1) \sin(y_1)\times \nonumber \\ &\cos(y_2)^2g_{10}^{y_1 y_1}g_{10}^{y_5 y_5} du \wedge dx \wedge dy_2 \wedge dy_4 \wedge dy_3-\cos(y_2) \sin(y_2)\cos(y_1)^2\times \nonumber  \\ &g_{10}^{y_2 y_2}g_{10}^{y_5 y_5} du \wedge dx \wedge dy_1 \wedge dy_3 \wedge dy_4)(2 \partial_u A_ x(u,t,z))+\mathcal{O}(A_x(u,t,z)^2),
\end{align}
\begin{align}
\frac{(F_5^{mag})^1_{tx}}{L(u)^4}&= -\sqrt{\det(g_{S_5})}\big(\sin(y_1)\cos(y_1)g_{10}^{y_1 y_1}g_{10}^{y_3 y_3} dt \wedge dx \wedge dy_2 \wedge dy_5 \wedge dy_4 +\nonumber \\ &  \cos(y_1)^2\sin(y_2)\cos(y_2)g_{10}^{y_2 y_2}g_{10}^{y_4 y_4} dt \wedge dx \wedge dy_1 \wedge dy_5 \wedge dy_3-\sin(y_1)\times \nonumber \\ &\cos(y_1)\sin(y_2)^2g_{10}^{y_1 y_1}g_{10}^{y_4 y_4} dt \wedge dx \wedge dy_2 \wedge dy_3 \wedge dy_5-\cos(y_1) \sin(y_1)\times \nonumber \\ &\cos(y_2)^2g_{10}^{y_1 y_1}g_{10}^{y_5 y_5} dt\wedge dx \wedge dy_2 \wedge dy_4 \wedge dy_3-\cos(y_2) \sin(y_2)\cos(y_1)^2\times \nonumber  \\ &g_{10}^{y_2 y_2}g_{10}^{y_5 y_5} dt \wedge dx \wedge dy_1 \wedge dy_3 \wedge dy_4)(2 \partial_t
 A_ x(u,t,z))+\mathcal{O}(A_x(u,t,z)^2),
\end{align}
\begin{align}
\frac{(F_5^{mag})^1_{zx}}{L(u)^4}&= -\sqrt{\det(g_{S_5})}\big(\sin(y_1)\cos(y_1)g_{10}^{y_1 y_1}g_{10}^{y_3 y_3} dz \wedge dx \wedge dy_2 \wedge dy_5 \wedge dy_4 +\nonumber \\ &  \cos(y_1)^2\sin(y_2)\cos(y_2)g_{10}^{y_2 y_2}g_{10}^{y_4 y_4} dz \wedge dx \wedge dy_1 \wedge dy_5 \wedge dy_3-\sin(y_1)\times \nonumber \\ &\cos(y_1)\sin(y_2)^2g_{10}^{y_1 y_1}g_{10}^{y_4 y_4} dz \wedge dx \wedge dy_2 \wedge dy_3 \wedge dy_5-\cos(y_1) \sin(y_1)\times \nonumber \\ &\cos(y_2)^2g_{10}^{y_1 y_1}g_{10}^{y_5 y_5} dz\wedge dx \wedge dy_2 \wedge dy_4 \wedge dy_3-\cos(y_2) \sin(y_2)\cos(y_1)^2\times \nonumber  \\ &g_{10}^{y_2 y_2}g_{10}^{y_5 y_5} dz \wedge dx \wedge dy_1 \wedge dy_3 \wedge dy_4)(2 \partial_z A_ x(u,t,z))+\mathcal{O}(A_x(u,t,z)^2),
 \label{eq:Fhodgeend2}
\end{align}
and 
\begin{equation}
F^{mag}=(F_5^{mag})^0+(F_5^{mag})^1_{tx}+(F_5^{mag})^1_{tx}+(F_5^{mag})^1_{ux}.
\end{equation}
Here $g_{S_5}$ denotes the metric of the five sphere, $g_{10}$ the $\gamma$-corrected metric of the entire manifold (\ref{eq:metricc}) and $g_5$ the $\gamma$-corrected metric of the internal AdS space. Now replace the $F_5^2$-term in the action with two times $(F^{mag})^2$ and insert the $\mathcal{O}(\gamma^0)$-solution of $F_5$ into the higher derivative term $\gamma W$ of the type IIB SUGRA action.  The result will be the new  action for $A_\mu$. Considering the prescription above we get together with 
\begin{equation}
\frac{1}{2 \times 5!}(F^{mag}_5)^2=\frac{8}{L(u)^{10}}+\frac{2 }{3L(u)^6}F_{\mu \nu}F^{\mu \nu}
\end{equation}
the following result for the part of the action depending on $A_\mu$, which doesn't contain higher derivative terms
\begin{equation}
-\frac{1}{2 \kappa_{10}}\int d^{10}x \sqrt{-g}\bigg(\frac{L(u)^2}{3}+\frac{2}{3L(u)^6}\bigg)F_{\mu \nu}F^{\mu \nu}.
\label{result12}
\end{equation}
The term $\frac{L(u)^2}{3}$ comes from the curvature scalar $R_{10}$.   Again we  only considered transverse fields $A_x$, respectively its Fourier transform $(A_x)_k$, with $k= (\omega, q)$.
The result for the $\gamma W$-part of the action given up to order $\mathcal{O}(A_x^2)$ is
 \begin{align}
&\frac{\gamma}{8 r_h^2}\int d^{10}x \sqrt{\det g_{10}} W=\gamma \text{vol}(S_5)\int \frac{d^4k}{(2\pi)^4} \bigg(A_W (A_x)_k'' (A_x)_{-k}+B_W (A_x)_k'(A_x)_{-k}' \nonumber\\&+ C_W (A_x)_k'(A_x)_{-k}+D_W (A_x)_k(A_x)_{-k}+E_W (A_x)_k'' (A_x)_{-k}''+F_W (A_x)_k''(A_x)_{-k}'\bigg)\nonumber \\ & + \mathcal{O}( \gamma^2)=: \gamma \text{vol}(S_5)\int \frac{d^4k}{(2 \pi)^4} \mathcal{L}_{\gamma}^1,
\end{align}
where the primes $'$ stand for $\partial_u$ and the functions $A_W,B_W,C_W,D_W,E_W,F_W$ are given by
\begin{align}
A_W= & \frac{4 u^5}{9}\bigg( 41 \hat{q}^2(1-u^2)-172 \hat{\omega}^2 \bigg) \label{estart}\\
B_W=& -\frac{2u^5}{9}\bigg( -803 u +1563 u^3 -216 \hat{q}^2 (1-u^2)-72 \hat{\omega}^2 \bigg)\\
C_W = & \frac{4 u^4}{9(1-u^2)}\bigg(\hat{q}^2(167-416 u^2 +249 u^4)-59 \hat{\omega}^2+511 u^2 \hat{\omega}^2  \bigg)\\
D_W=&\frac{2 u^3}{9 (1-u^2)^2}\bigg(-90 \hat{q}^4 u (1-u^2)^2-\hat{\omega}^2(270-441u^2+99 u^4+208 u \hat{\omega}^2)+\hat{q}^2\nonumber \\ & (1-u^2)(162 -315 u^2 +153 u^4 -134 u \hat{\omega}^2)  \bigg)
\end{align}
\begin{align}
&E_W=  -\frac{416}{9}u^6(1-u^2)^2\\
&F_W= -\frac{20 u^5}{9} \bigg(37-150 u^2+113 u^4 \bigg)\label{eend}.
\end{align}
We already used  the definitions $\hat{\omega}= \frac{\omega}{2 \pi T}= \frac{\omega}{2 r_h}+\mathcal{O}(\gamma)$ and $\hat{q}= \frac{q}{2 \pi T}= \frac{q}{2 r_h}+\mathcal{O}(\gamma)$
here.  Together with (\ref{result12}) equations (\ref{estart})-(\ref{eend}) explicitly give the $\mathcal{O}(\gamma)$-Lagrangian for $(A_x)_k$ up to second order in $(A_x)_k$ .
\begin{align}\mathcal{L}=&\frac{1}{2}\bigg((A_x)_k(A_x)_{-k}\Big(\frac{\tilde{q}^2(1-u^2)-\tilde{\omega}^2}{u(1-u^2)}+\gamma \frac{5 u }{8 (1-u^2)}\Big( -10 \tilde{q}^2u^2-197 \tilde{q}^2u^4 +207 \tilde{q}^2 u^6 \nonumber \\ &-130 \tilde{\omega}^2-120 u^2 \tilde{\omega}^2 +274 u^4 \tilde{\omega}^2\Big)  \Big)+(A_x)_k'(A_x)_{-k}'(1-u^2)\Big(1+\gamma \frac{5}{16}\Big( -260 u^2- \nonumber \\ &235 u^4+553 u^6\Big) \Big)\bigg)+\mathcal{L}_{\gamma}^1,
\end{align}
with $\tilde{\omega}= \frac{\omega}{2 r_h}$ and $\tilde{q}= \frac{q}{2 r_h}$.\\ \indent 
In the next step we derive the $\gamma$-corrected EoM for our gauge field $(A_x)_k$ by  varying the action with respect to $(A_x)_k$. We do not want to focus on boundary terms here but merely on the resulting EoM for $(A_x)_k$. This simple exercise gives 
\begin{equation}
2 \left(u^2-1\right) (A_x)_k''+4 u (A_x)_k'+\frac{2 (A_x)_k \left(\tilde{q}^2 \left(u^2-1\right)+\tilde{w}^2\right)}{u \left(u^2-1\right)}-\gamma H((A_x)_k)=\mathcal{O}(\gamma^2)
\end{equation}
with
\begin{align}
 H((A_x)_k)=&\frac{u}{72 (u^2-1)^2}\Bigg((u^2-1)^2 \bigg(u (A_x)_k'' (-8576 \hat{q}^2 u^5+128 u^3 (67 \hat{q}^2+208 \hat{\omega}^2)\nonumber \\ &+1398243 u^6-1740092 u^4+459685  u^2-11700)+4 (A_x)_k'
   (-15008 \hat{q}^2 u^5\nonumber \\ &+160 u^3 (67 \hat{q}^2+208 \hat{\omega}^2)+401046 u^6-373722 u^4+60325 u^2-5850)+\nonumber \\&13312 (u^2-1) u^4 (u (u^2-1) (A_x)_k''''+4
   (5 u^2-3)(A_x)_k''')\bigg)+2 (A_x)_k (2880 \nonumber  \\ & \hat{q}^4u^3 (u^2-1)^2+\hat{q}^2 u^2 (u^2-1) (21507  u^4-31105 u^2-4288 u
   w^2+9598)+\nonumber  \\ & 2 \hat{\omega}^2(30085 u^6-75057 u^4+3328 u^3 \hat{\omega}^2+55359  u^2+2925))\Bigg).
   \label{eq:H}
\end{align}
Exploiting that we have 
\begin{equation}
(A_x)_k''+\frac{2 u (A_x)_k'}{ \left(u^2-1\right) }+\frac{ (A_x)_k \left(\hat{q}^2 \left(u^2-1\right)+\hat{w}^2\right)}{u \left(u^2-1\right)^2}=\mathcal{O}(\gamma)
\end{equation}
reduces (\ref{eq:H}) to
\begin{align}
\gamma H((A_x)_k)=&\frac{u \gamma }{72(1-u^2)}\bigg((A_x)_k (-27648 \hat{q}^4 u^3 (u^2-1)+\hat{q}^2 (370501 u^6-666170 u^4+\nonumber\\ &307369 u^2-11700)+9\hat{\omega}^2 (5951 u^4-9081 u^2+2600) )-10
   (u^2-1)^2 \nonumber \\ &(-17600 \hat{q}^2 u^3+8493 u^4-19450 u^2+2340) (A_x)_k'\bigg)+\mathcal{O}(\gamma^2).
\end{align}
To simplify this further we define 
\begin{equation}
\Sigma(u)=\frac{5 \gamma  \left(-7040 q^2 u^5+2831 u^6-9725 u^4+2340 u^2\right)}{288 \sqrt{1-u^2}}+\frac{1}{\sqrt{1-u^2}},
\end{equation}
so that with 
\begin{equation}
\Psi= (A_x)_k/ \Sigma(u)
\end{equation}
 we end up with the following EoM
\begin{align}
0=& \Psi''+\Psi\bigg(\frac{u-\hat{q}^2(1-u^2)+\hat{\omega}^2}{u (1-u^2)^2} -\frac{\gamma }{144 u (1-u^2)}\bigg(-27648 \hat{q}^4 u^5+\hat{q}^2 (-157499 u^6+\nonumber \\ &56331 u^4+11700 u^2+4770)+297255 u^7-698575u^5+53559 u^4 \hat{\omega}^2+326850 u^3\nonumber \\ &-28170 u^2 \hat{\omega}^2-11700 u-4770 \hat{\omega}^2 \bigg)\bigg),
\label{eq:eomfinal}
 \end{align} where we already used the $\gamma$ corrected relation between the temperature and $r_h$
 \begin{equation}
 r_h= \pi T \Big(1-\frac{265}{16} \gamma \Big).
 \end{equation}
From this differential equation one obtains identical $\gamma$-corrections for the conductivity, the photoemission rate and the QNM spectrum for all values of $\hat{q}$ considered. We want to highlight that this is firstly almost certainly not a coincidence and secondly comes very unexpectedly. On the one hand this coupling corrected differential equation  (\ref{eq:eomfinal}) should be taken with a grain of salt, since unlike (\ref{finaleq1}) and (\ref{finaleq2}) it doesn't follow mathematically, but by intuitively extending a calculational prescription into a regime, where it actually shouldn't hold anymore. On the other hand, since especially the coupling corrections to the QNM are identical in both our calculations, one could argue that it isn't a surprise that other quantities coincide with what we found previously. This is because the QNM govern huge parts of the behaviour of our system. 
\section{Discussion} 
In this paper we rederived the finite coupling correction to the EoM for gauge fields and corrected several mistakes found in the literature. We  have computed finite coupling corrections to the photoemission rate, the electrical conductivity, the QNM spectrum for different momenta and (off equilibrium) spectral density of a $\mathcal{N}=4$ SYM plasma. We analyzed the behaviour of  QNMs for realistic values of $\lambda_{\text{'t Hooft}}$ using the partial resummation technique starting from the full $\mathcal{O}(\alpha'^3)$-corrections to the SUGRA action. We saw that in both the large and small energy limit the corrections to the spectral function respectively the photoemission rate behave as expected from (perturbative) weak coupling calculations
  \cite{CaronHuot:2006te}. Interestingly we found that the term in the EoM for coupling corrected gauge fields (\ref{finaleq1}) governing the large $\omega$ behaviour, whose existence is crucial for the right behaviour of the photoemission rate in this region is precisely the same as in the spin-$2$ channel. \\ \indent The resummation technique, which in principle is an approximation using the assumption that the correction terms to (\ref{finaleq1}), (\ref{finaleq2}) of order higher than $\mathcal{O}(\alpha'^3)$ are small, whereas the first order correction approximates real physics by assuming that the higher order corrections to the quantities of interest themselves  are small, can also be applied in an analogous way to the conductivity. For $\lambda=11.3 \approx 4 \pi N \alpha_s|_{N=3, \alpha_s=0.3}$  we obtain a resummed value  of \begin{equation}
  \sigma= 0.29082 e^2 T. 
  \label{rescon}
  \end{equation}
  This can be compared to results of hot QCD lattice calculations. For temperatures above $T_c$ the authors of \cite{Aarts} found $\sigma \approx e^2 T (0.4 \pm 0.1)$. More recently this could be improved to $\sigma\approx e^2 T(0.31 \pm 0.05) $ for $T > 1.75 T_c $, see figure $10$ in \cite{Aarts2}. Without any coupling corrections the conductivity is given by \begin{equation}
 \sigma_{\infty} = \frac{9}{16  \pi}e^2 T \approx  0.179 e^2 T.
  \end{equation}
 In conclusion the coupling corrected and resummed result comes noticeably closer to hot-QCD lattice results. 
  In the last part we  note a surprising observation. If we naively extend the prescription valid in   $\mathcal{O}(\gamma^0)$ to the  $\mathcal{O}(\gamma^1)$ case, which leads to a quite different EoM for $A_x$ than our strict derivation in section \ref{sec3}, we still obtain the same corrections to the QNM-spectra, the conductivity and the photoemission rate in both the large and the small $\omega$ limit as in section \ref{sec:results}. However, the results for the off-equilibrium spectral density were different. It certainly would be interesting to understand why one obtains correct answers for the equilibrium observables we calculated, because, although still tedious, the calculation is significantly easier than the one in section \ref{sec3}. 
\begin{table}[t]
%
\begin{center}

\begin{tabular}{|c|c|c|c|}
\hline
    Quantity &
    $\mathcal{O}(\gamma^0)$ &
    $\mathcal{O}(\gamma^1)$ &
    Reference
\\ \hline
    $s \, (\tfrac 12 \pi^2 \Nc^2 \, T^3)^{-1}$ &
    $1$ &
    $15 \, \gamma$ &
    \cite{Gubser}
\\ \hline
    $\eta \, (\tfrac 18 \pi \Nc^2 \, T^3)^{-1}$ &
    $1$ &
    $135 \, \gamma$ &
    \cite{Buchel:2008sh}
\\ \hline
    $4\pi \, \eta/s$ &
    $1$ &
    $120 \, \gamma$ &
    \cite{Buchel:2008sh}
\\ \hline
    $\sigma \, (\tfrac 14 \alpha_{\rm EM} N^2 \, T)^{-1 }$ &
    1 &
    $125\, \gamma $&
    This work
\\ \hline
    ${\omega}^{\rm shear}_2(q=0) \, (2 \pi T)^{-1}$  &
    $2.585-2.382 \, i $ &
    $(1.029+0.957 \, i) \, 10^4 \, \gamma$  &
    \cite{Stricker:2013lma}
\\ \hline
   ${\omega}^{\text{EM}}_2(q=0) \, (2 \pi T)^{-1}$  &
    $2-2 \, i $ &
    $(4.896 + 0.495\, i) \, 10^3 \, \gamma$  &
    This work
\\ \hline
\end{tabular}
\end{center}
\caption
    {%
    \small
    A collection of results  for the zeroth and first order terms in the expansion of
    various thermal observables in powers of
    $\gamma=\frac 18\, \zeta(3)\lambda^{-3/2}$.
    Results are shown for the entropy density $s$,
    shear viscosity $\eta$,
    viscosity to entropy density ratio $\eta/s$,
    electrical conductivity $\sigma$
    and the second quasinormal mode frequencies,
    $\omega^{\rm EM}_2$ and $\omega^{\rm shear}_2$,
    at zero wave vector, for the electromagnetic current
    and shear channel of the stress-energy correlator,
    respectively.
   \label{table:1}
    }
\end{table}
\FloatBarrier
\section{Appendix}
 \label{Appendix}
\begin{align}
\bigg(\frac{\delta \mathcal{W}}{\delta F_5}\bigg)_{xy_1y_2y_4y_5}=&\gamma \frac{\cos(y_1)^3 \sin(y_1) \sin(y_2) \cos(y_2)}{6 \sqrt{3}}\big(  (-117 \frac{u^5}{(1 - u^2)}\partial_t^2 A_x + (468 u^6 - \nonumber\\ &468 u^8)\partial_u^2 A_x + 
   83 u^5 \partial_z^2 A_x + (4312 u^5 - 5248 u^7)\partial_u A_x)\big)+\mathcal{O}(A_x^2)
\end{align}
\begin{align}
\bigg(\frac{\delta \mathcal{W}}{\delta F_5}\bigg)_{x zy_2y_4y_5}=&-\gamma \frac{(\cos(y_1)^4  \sin(y_2) \cos(y_2))}{6 \sqrt{3}}\big((415 u^6 - 415 u^8) \partial_u^2\partial_z A_x +\frac{415 u^5}{4(-1 + u^2) }\nonumber \\& 
   \partial_z\partial_t^2 A_x   - 
 \frac{261 u^5}{4} \partial_z^3 A_x+ (3220 u^5 - 4050 u^7) \partial_u\partial_z A_x + \nonumber \\& (2181u^4 - 3216 u^6) \partial_z A_x \big)+\mathcal{O}(A_x^2)
\end{align}
\begin{align}
\bigg(\frac{\delta \mathcal{W}}{\delta F_5}\bigg)_{u x y_2y_4y_5}=&\gamma \frac{(\cos(y_1)^4  \sin(y_2) \cos(y_2))}{6 \sqrt{3}}\big((-733 u^6 + 733 u^8) \partial_u^3A_x +\frac{733 u^5}{4 (1 - u^2)} \nonumber \\& \partial_u \partial_t^2 A_x - 
\frac{257 u^5}{4} \partial_z^2 \partial_u A_x+ (-4398 u^5 + 7330 u^7) \partial_u^2 A_x +\nonumber \\& \frac{3117 u^4 - 1651 u^6}{4(1-u^2)^2}   \partial_t^2 A_x- 
 1145 u^4 \partial_z^2 A_x + (-2056 u^4 + 12162 u^6) \partial_u A_x \big)\nonumber \\& +\mathcal{O}(A_x^2)
\end{align}
\begin{align}
\bigg(\frac{\delta \mathcal{W}}{\delta F_5}\bigg)_{t x y_2y_4y_5}=&-\gamma \frac{(\cos(y_1)^4  \sin(y_2) \cos(y_2))}{6 \sqrt{3}}\big((733 u^6 - 733 u^8) \partial_u^2 \partial_t A_x + 
 \frac{257 u^5}{4}\partial_z^2 \partial_t A_x\nonumber \\&+ \frac{733 u^5}{ 4 (-1 + u^2)} \partial_t^3 A_x + (548 u^5 - 2014 u^7) \partial_u\partial_t A_x + (u^4 (-609 + 912 u^2) ) \nonumber \\& \partial_t A_x \big)+\mathcal{O}(A_x^2)
\end{align}
\section*{Acknowledgements}
We thank Christian Ecker, Martin Schvellinger, Aleksi Vuorinen and  Laurence Yaffe for useful discussions and remarks. The work of SW was supported by the Elite Network of Bavaria. He  thanks the University of Helsinki for their support and hospitality during portions of this work.


\begin{thebibliography}{99}
\bibitem{Hassanain:2011fn}
    B.~Hassanain and M.~Schvellinger,
    {\it Plasma conductivity at finite coupling,}
    \jhep {1201}{2012}{114},
    \arXivid{1108.6306}.
    
\bibitem{Hassanain:2012uj}
  B.~Hassanain and M.~Schvellinger,
  {\it Plasma photoemission from string theory,}
  \jhep {1212}{2012}{095},
  \arXivid{1209.0427} [hep-th].
  
  \bibitem{Hassanain:2011ce}
    B.~Hassanain and M.~Schvellinger,
    {\it Diagnostics of plasma photoemission at strong coupling,}
    \prd {85}{2012}{086007},
    \arXivid{1110.0526} [hep-th].
    
\bibitem{Pawelczyk:1998pb}
  J.~Pawelczyk and S.~Theisen,
  {\it $AdS_5 \times S^5$ black hole metric at $O(\alpha'^3)$,}
  \jhep {9809}{1998}{010},
  \hepth{9808126}.
  
\bibitem{Paulos:2008tn}
  M.~F.~Paulos,
  {\it Higher derivative terms including the Ramond-Ramond five-form,}
  \jhep {0810}{2008}{047},
  \arXivid{0804.0763} [hep-th].
  
  \bibitem{Son:2002}
    D.~T.~Son, A.~O.~Starinets,
    {\it Minkowski-space correlators in AdS/CFT correspondence:
    recipe and applications,}
    \jhep {0209}{2002}{042},
    \hepth{0205051}.
    
\bibitem{CaronHuot:2006te}
  S.~Caron-Huot, P.~Kovtun, G.~D.~Moore, A.~Starinets and L.~G.~Yaffe,
  {\it Photon and dilepton production in supersymmetric Yang-Mills plasma,}
  \jhep {0612}{2006}{015},
  \hepth{0607237}.
  
  \bibitem{Buchel:2008sh} 
  A.~Buchel,
  {\it Resolving disagreement for $\eta/s$ in a CFT plasma at finite coupling,}
  \npb {803}{2008}{166},
  \arXivid{0805.2683} [hep-th].
  
  \bibitem{Vuorinen:2013}
    D.~Steineder, S.~A.~Stricker, A.~Vuorinen,
    {\it Probing the pattern of holographic thermalization with photons,}
    \jhep {1307}{2013}{014},
    \arXivid{1304.3404}.
    
    
    \bibitem{Chamblin}
    A.~Chamblin, R.~Emparan, C.~V.~Johnson, R.~C.~Myers,
    {\it Charged AdS Black Holes and Catastrophic Holography}
    \arXivid{hep-th/9902170}.
    
        \bibitem{ourpaper}
    S.~Waeber., A.~Sch\"afer., A.~Vuorinen., L.~G.~Yaffe.,
    {\it Finite coupling corrections to holographic predictions for hot QCD}
    \jhep {1511}{2015}{087}
    \arXivid{1509.02983}.
    
    \bibitem{Gubser}
    S.~S.~Gubser, I.~R.~Klebanov, A.~A.~Tseytlin 
    {\it
Coupling Constant Dependence in the Thermodynamics of $\mathcal{N}=4$ Supersymmetric Yang-Mills Theory}
    \arXivid{9805156} [hep-th].
   \bibitem{Witten}
     D.~J.~Gross and E.~Witten, 
    {\it
Superstring modifications of Einstein's equations}
    \npb {277}{1986}{1}
    
     \bibitem{Aarts}
     G.~Aarts, C.~Allton, J.~Foley, S.~Hands, S.~Kim, 
    {\it
 Spectral functions at small energies and the electrical conductivity in hot, quenched lattice QCD}
    \arXivid{0703008} [hep-th].
\bibitem{Aarts2}
     G.~Aarts, C.~Allton, A.~Amato, P.~Giudice, S.~Hands, J.~Skullerud, 
    {\it
Electrical conductivity and charge diffusion in thermal QCD from the lattice}
    \arXivid{1412.6411} [hep-th].
    
\bibitem{Stricker:2013lma}
  S.~A.~Stricker,
  {\it Holographic thermalization in $\Nfour$ super-Yang-Mills theory
  at finite coupling,}
  \epjc {74}{2014}{2727},
  \arXivid{1307.2736}.
\bibitem{Buchel:2005}
    A.~Buchel, J.~T.~Liu, A.~O.~Starinets,
    {\it Coupling constant dependence of the shear viscosity in
    $\Nfour$ supersymmetric Yang-Mills theory,}
    \npb {707}{2005}{56-68},
    \hepth{0406264}.
    
    \bibitem{Peeters}
    K.~Peeters, A.~Westerberg,
    {\it The Ramond-Ramond sector of string theory beyond leading order}
    \arXivid{hep-th/0307298}.
\bibitem{Skenderis}
   S.~de Haro, A.~Sinkovics, K.~Skenderis,
    {\it On a supersymmetric completion of the $R^4$ term in IIB supergravity}
    \arXivid{hep-th/0210080}.
\bibitem{Skenderis2}
   S.~de Haro, A.~Sinkovics, K.~Skenderis,
    {\it On alpha'-corrections to D-brane solutions}
    \arXivid{hep-th/0302136}.
\end{thebibliography}
 \end{document}